\documentclass[12pt]{iopart}
\usepackage{latexsym}

\def\cH{{\cal H}}

\def\tr{{\rm tr}}
\def\ket#1{\mid~\!\!\!{#1}~\!\!\rangle}
\def\bra#1{\langle~\!\!{#1}~\!\!\!\mid}
\def\IF{if and only if }
\def\QM{quantum mechanics }
\def\qm{quantum mechanics}
\def\QMl{quantum-mechanical }

\def\cR{{\cal R}}
\def\${\enskip$}
\def\M{measurement }
\def\m{measurement}
\def\Q{quantum }

\def\I{interpretation }
\def\i{interpretation}

\def\D{Definition }

\def\L{Lemma }

\def\P{Proposition }

\def\C{Corollary }

\def\R{Remark }

\def\T{Theorem }

\begin{document}\jl{1}

\title[Subsystem expansion]{Bipartite Entanglement\\ Review of Subsystem-Basis Expansions\\  and Correlation Operators in It\\
}

\author{F Herbut}

\address{Serbian Academy of
Sciences and Arts, Knez Mihajlova 35,
11000 Belgrade, Serbia\\ E-mail:
fedorh@sanu.ac.rs}

\date{\today}

\begin{abstract}
The present review presents the authors previous results on the topic from the title in a new light.
Most of the previous results were obtained using the techniques of antilinear Hilbert-Schmidt mappings of one Hilbert pace onto another, which is unknown and unused in the literature. This, naturally, diminished the impact of the results. In this article the results are derived anew with standard techniques. The topics listed at the end of the Introduction, are expounded in 9 theorems, 5 propositions etc. Partial scalar product and partial trace methods are used throughout. Further relevant research articles that are not reproduced in this review, are sketched in the Concluding remarks.
\end{abstract}


\maketitle

\rm

\section{Introduction}

\noindent
The aim of this study is to present a
detailed and elaborated exposition of the
subject in the title with almost all claims
proved by arguments that mostly do not
coincide with those in the original
articles. Namely, since much work has
been done so far in the
Belgrade school, the present-day views
are more mature and hence they differ
from the originally perceived ones, and make possible simpler proofs. This
fact alone should
justify writing most of this review as if
it were done for the first time. There is
also the additional fact that many
results of the research have been previously
presented in the formalism in which bipartite state vectors are written as  antilinear
Hilbert-Schmidt operators mapping one subsystem state space into the other (cf
\cite{JMPstructure}), which is not well
known, and it is very rarely used. Eventually, this approach has been found replaceable by standard basis-independent treatment. The basic aim is an in-depth study of the Schmidt decomposition. Its various forms are presented with its underlying foundations in three layers.\\

We assume that a completely arbitrary
{\bf bipartite state vector}
\$\ket{\Psi}_{AB}\$ is given. It is an
arbitrary normalized vector in
\$\cH_A\otimes\cH_B,\$ where the factor
spaces are finite- or
infinite-dimensional complex separable
Hilbert spaces, the state spaces of the subsystems A and B. The statements are, as
a rule, asymmetric in the roles of the
two factor spaces. But, as it is well
known, for every general asymmetric
statement, also its symmetric
counterpart, obtained by exchanging the
roles of subsystems A and B, is valid.

Having in
mind local, i. e., subsystem \m , we
choose arbitrarily that it is performed on subsystem
B. (That this choice is practical for presentation will be obvious in relation (30) below and further.) We call  subsystem B the 'nearby'
one, and the opposite subsystem A, which
is not affected dynamically by the local
\m , we call 'distant'. This is not a
synonym for 'far away'. But the
suggestion of the latter may help to
picture the lack of dynamical
influence on subsystem A.

The basic mathematical tool in the
analysis are the partial scalar product
(elaborated in Appendix A) and the rules of the partial trace (presented and proved in Appendix B).

Hermitian operators, i. e., observables and subsystem state operators
(density operators) will be given, unless otherwise stated,
in their so-called {\it 'unique' spectral
forms}, which are defined by lack of
repetition in the eigenvalues. For
instance, \$O=\sum_ko_kP_k,\enskip k\not=
k'\enskip\Rightarrow\enskip o_k\not=
o_{k'}\$, where "$\Rightarrow$" denotes
logical implication. Then \$P_k\$ is said
to be the eigen-projector of \$O\$ that
corresponds to the eigenvalue \$o_k\$,
and its range \$\cR(P_k)\$ is the
corresponding eigen-subspace. We consider only Hermitian operators that have a purely discrete spectrum. We call them discrete operators.

Vectors that are not necessarily of norm
one are written overlined throughout.
Besides, when a number multiplies from
the left a vector or an operator, the
multiplication symbol \$\times\$ is put
between them for clarity. One should
keep in mind the convention that if a
term in a sum has two or more factors and
the first is zero, the rest need not be
defined; it is understood that the entire
term is zero. In tensor products of
vectors we put only occasionally the
tensor multiplication sign '$\otimes$'
when more clarity is required. By "basis" we mean a complete ortho-normal set of elements throughout.\\

The reader will not find, hopefully, the
abundant use of mathematical
structure (theorems, propositions,
lemmata, corollaries, remarks and
definitions) annoying. They are
important for the many cross-references
in the present paper, as well as for references in future articles. Besides, they reveal the logical status of the claim they contain.\\

The arrangement of the exposition in
sections and subsections goes as
follows.

2 Expansion in a subsystem basis

3 Schmidt decomposition

4 Correlated Schmidt decomposition

5 Twin-correlated Schmidt decomposition

6 Distant measurement and EPR states

6.1 Distant \M

6.2 EPR states

6.3 Schroedinger's steering

7 Concluding remarks

AppA  Partial scalar product

AppB  The partial-trace rules

\section{Expansion in a subsystem basis}

The natural framework for
Schmidt (or biorthogonal) decomposition
is decomposition in a factor-space basis, or, as we shall call it, expansion in a subsystem basis.\\

{\bf Theorem 1. A)} Let \$\ket{\Psi}_{AB}\enskip\Big(\in (\cH_A\otimes\cH_B)\Big)\$ be any bipartite state vector. Let further \$\{\ket{n}_B:\forall n\}\$ be an arbitrary basis in the state space \$\cH_B\$. Then there exists a {\bf unique expansion in the subsystem basis}
$$\ket{\Psi}_{AB}= \sum_n\overline{\ket{n}}_A
\ket{n}_B.\eqno{(1a)}$$

The 'expansion coefficients' \$\overline{\ket{n}}_A\$ have the form
$$\forall n:\quad\overline{\ket{n}}_A=\sum_m(\bra{m}_A
\bra{n}_B\ket{\Psi}_{AB})\times\ket{m}_A,\eqno{(1b)}$$
where \$\{\ket{m}_A:\forall m\}\$ is an arbitrary basis in \$\cH_A\$. The 'expansion coefficients' \$\overline{\ket{n}}_A\$ in (1a) are elements in \$\cH_A\$, and they are not necessarily of norm one. They depend {\bf only} on \$\ket{\Psi}_{AB}\$ and the
corresponding basis elements
\$\ket{n}_B\$, and not on the choice of the rest of basis elements in the basis \$\{\ket{n'}_B:\forall n'\}\$.

The sums in (1a) and (1b), if infinite, are absolutely convergent, and one has
$$||\ket{\Psi}_{AB}||^2=\sum_n||\overline{\ket{n}}_A||^2,
\eqno{(1c)}$$ as well as $$\forall n:\quad ||\overline{\ket{n}}_A||^2=\sum_m |\bra{m}_A\bra{n}_B\ket{\Psi}_{AB}|^2.\eqno{(1d)}$$
In case of infinity, each of the sums is an absolutely convergent series as 'inherited' from the absolutely convergent series $$\ket{\Psi}_{AB}= \sum_{mn}[(\bra{m}_A
\bra{n}_B\ket{\Psi}_{AB})\times\ket{m}_A
\ket{n}_B].\eqno{(1e)}$$

Further, one can suitably write \$\forall n:\enskip \overline{\ket{n}}_A=||\overline{\ket{n}}_A||\times \ket{n}_A\$ (definition of the norm -one elements \$\{\ket{n}_A:\forall n\}\$), and replace these in (1a). Relation (1a) then becomes $$\ket{\Psi}_{AB}=\sum_n||\overline{\ket{n}}_A||\times \ket{n}_A\ket{n}_B,\eqno{(1f)}$$ This is an expansion  in the ON set of elements \$\{\ket{n}_A\ket{n}_B:\forall n\}\$ in \$\cH_A\otimes\cH_B\$. Actually, it is 'normal' in both factors, but orthogonal, in general, only in the second one. Some norm-one elements \$\ket{n}_A\$ may not exist, when \$\overline{\ket{n}}_A=0\$ (depending on \$\ket{\Psi}_{AB}\$).

{\bf B)} The expansion coefficients can be evaluated utilizing the {\bf partial scalar product}   $$\forall n:\quad\overline{\ket{n}}_A
=\Big(\bra{n}_B\ket{\Psi}_{AB}\Big)_A.\eqno{(1g)}$$\\

{\bf Proof A)} is straightforward, but, on account of the importance of the theorem (see end of the section), it is presented as easy reading.

Let \$\{\ket{m}_A:\forall m\}\$ be an arbitrary basis in \$\cH_A\$. Then one can perform the expansion (1e). As it is well known, if the double-sum is infinite, the series is absolutely convergent allowing any change of order in which the terms are written (any permutation). Hence we can group together all terms around each \$\ket{n}_B\$ tensor factor and rewrite (1e) as  $$\ket{\Psi}_{AB}= \sum_n\Big(\sum_m(\bra{m}_A
\bra{n}_B\ket{\Psi}_{AB})\times\ket{m}_A\Big)
\ket{n}_B.$$ Thus one obtains (1a) and (1b).

In this way we have established that the claimed expansion exists. Now we show that the 'expansion coefficients' \$\overline{\ket{n}_A}\$ in (1a) do not depend on the choice of the basis \$\{\ket{m}_A:\forall m\}\$. Let  \$\{\ket{k}_A:\forall k\}\$ be any other basis in \$\cH_A\$, and let \$\forall m:\enskip \ket{m}_A=\sum_kU_{m,k}\ket{k}_A\$ be the unitary transition matrix. Then, starting with the 'expansion coefficient' evaluated in the first basis, we find out its form in the second basis:
$$\overline{\ket{n}_A}=\sum_m\Big((\bra{m}_A
\bra{n}_B\ket{\Psi}_{AB})\times\ket{m}_A\Big)=$$
$$\sum_k\sum_{k'}\sum_mU_{m,k}^*U_{m,k'}(\bra{k}_A
\bra{n}_B\ket{\Psi}_{AB})\times\ket{k}_A.$$ Since \$\sum_mU_{m,k}^*U_{m,k'}=\delta_{k,k'}\$ is valid for the unitary transition matrix elements, one is further led to
$$\overline{\ket{n}_A}=\sum_k(\bra{k}_A
\bra{n}_B\ket{\Psi}_{AB})\times\ket{k}_A.$$
Obviously, if the 'expansion coefficient' were evaluated in the other basis, it would give the same element of \$\cH_A\$. The additional claims in (A) are obvious.

{\bf B)} Proof of (1g) is given in Appendix A, where the partial scalar product is defined in 'three and a half ways'; one of them consisting precisely in equating RHS(1b) and RHS(1g).)\hfill $\Box$\\

{\bf Corollary 1.} If the nearby state is pure, i. e., a state vector, e. g.
\$\ket{\bar n}_B\$, then also the distant
state is necessarily pure, but it can be  arbitrary (depending on \$\ket{\Psi}_{AB}\$).\\

{\bf Proof.} By assumption
\$\rho_B\equiv\tr_A
\Big(\ket{\Psi}_{AB}\bra{\Psi}_{AB}\Big)=
\ket{\bar n}_B\bra{\bar n}_B\$. Choosing
a  nearby-subsystem basis
\$\{\ket{n}_B:\forall n\}\$ so that it
contains \$\ket{\bar n}_B\$, one obtains
\$\bra{n}_B\rho_B\ket{n'}_B=\delta_{n,\bar
n}\delta_{n',\bar n}\ket{\bar
n}_B\bra{\bar n}_B\$.

On the other hand, expansion (1a) implies
$$\bra{n}_B\rho_B\ket{n'}_B=\bra{n}_B\tr_A
\Big(\ket{\Psi}_{AB}\bra{\Psi}_{AB}\Big)\ket{n'}_B=
\overline{\ket{n}_A}\,\overline{\bra{n'}_A}.$$
Altogether,
$$\overline{\ket{n}_A}\overline{\bra{n'}_A}=
\delta_{n,\bar n}\delta_{n',\bar
n}\ket{\bar n}_B\bra{\bar n}_B,$$ i. e.,
\$\forall
n:\enskip\overline{\ket{n}_A}=\delta_{n,\bar
n}\overline{\ket{\bar n}_A}\$. Hence,
\$\ket{\Psi}_{AB}=\ket{\bar
n}_A\ket{\bar n}_B\$ (\$\overline{\ket{\bar n}_A}\$ is of norm one because so are \$\ket{\Psi}_{AB}\$ and \$\ket{\bar n}_B\$).\hfill $\Box$\\

As an alternative proof of \C 1 one may
consider the canonical Schmidt
decomposition (cf \D 3 and relation (5)
together with (6a,b) below). Then the
claim in \C 1 is obvious, but the burden
of the proof lies on \T 3.\\

We define a term known in the
literature.

{\bf Definition 1.} If \$\rho\$ is an
arbitrary mixed state (density operator
that is not a rewritten state vector) of
a \Q system in the state space (Hilbert
space) \$\cH\$, then one can
isomorphically map \$\cH\$ onto the
subsystem state space \$\cH_A\$ of a
bipartite \Q system the state of which is
in \$\cH_A\otimes\cH_B\$, and find a
state vector \$\ket{\Psi}_{AB}\$ such
that its first-subsystem state operator (reduced
density operator) \$\rho_A\$ is
isomorphic to the initially given
\$\rho\$. This procedure is called {\bf
purification}.\\

{\bf Theorem 2. On purification.} Any
mixed state \$\rho\$ can be {\bf
purified} if it is written as any mixture
$$\rho=\sum_n\overline{\ket{n}}\,\overline{\bra{n}}\eqno{(1h)}$$
by writing down a bipartite state vector \$\ket{\Psi}_{AB}\$
in the form (1a) with any basis in
\$\cH_B\$, denoted as \$\{\ket{n}_B:\forall n\}\$, with expansion coefficients \$\overline{\ket{n}}\$ given by (1c) with added index A. The subsystem state operator (reduced
density operator) \$\rho_A\$ is then
isomorphic to \$\rho$.\\

{\bf Proof.} Evaluating \$\rho_A\equiv\tr_B\ket{\Psi}_{AB}\bra{\Psi}_{AB}\$ and keeping in mind that \$tr(\ket{n}_A\bra{n'}_A)=
\bra{n'}_A\ket{n}_A=\delta_{n,n'}\$, one obtains \$\rho_A=\sum_n\overline{\ket{n}}_A\overline{\bra{n}}_A\$.\hfill $\Box$\\

To understand the importance of subsystem-basis decomposition (1a), one must realize that
expansion (1a) is a {\it cross-road}. A number of different paths lead from it:

(i) Definition of the {\it partial scalar product}. Von Neumann in his seminal book \cite{VNeum}, in which he gave the mathematical grounding of \QM in case of infinite-dimensional state spaces, did not encompass partial scalar product and partial trace. Therefore, a careful mathematical exposition of these concepts is given, together with the basic properties, in Appendices A, B and C.

(ii) The expansion at issue leads to {\it purification} (cf Theorem 2 and relation (1d) above).

(iii) It is the framework for {\it Schmidt decomposition} (see section 3 and further).

(iv) Remark 5 and relation (12) below open the way for a more fruitful application of (1a), particularly for Schr\"o dinger's important concept of {\it steering} (cf subsection 6.3 below).

(v) Expansion (1a) gives a new angle on the concept of {\it erasure} (cf Remark 22 below).

(vi) A theory of {\it preparation} in \Q experiments can be based on (1a): If the preparator is sybsystem B, and the object on which the experiment is conducted is subsystem A, and if \$\ket{\Psi}_{AB}\$ is the state after interaction, then \$\ket{n}_B\$ is the state of the preparator that the experimenter 'sees' at the end of the preparation, and simultaneously \$\ket{n}_A\$ is then the state of the experimental object (at the beginning of the experiment). This will be elaborated in future work.

(vii) Expansion (1a) can play a crucial role in Everett's {\it relative-states interpretation} of \qm : The state \$\ket{n}_A\$ is the relative state of subsystem A with respect to the state \$\ket{n}_B\$ of subsystem B in the composite-system state \$\ket{\Psi}_{AB}\$. A detailed discussion of this and its ramifications is left for future work.

Subsystem-basis expansion (1a), and the enumerated paths (i) and (iv)-(vii) that lead away from it were not analyzed in previous work. This material is {\it new} in this article.\\

\section{Schmidt decomposition}

Now we define Schmidt (or biorthogonal)
decomposition. It is well known and much
used in the literature.\\

{\bf Definition 2.} If besides the basis
elements \$\ket{n}_B\$ also the expansion
coefficients \$\overline{\ket{n}_A}\$ are
{\it orthogonal} in expansion (1a), then
one speaks of a {\bf Schmidt or
biorthogonal decomposition}. It is
usually written in terms of subsystem
state vectors \$\{\ket{n}_A:\forall n\}\$
that are not only orthogonal, but also
normalized:
$$\ket{\Psi}_{AB}=\sum_n\alpha_n\ket{n}_A
\ket{n}_B,\eqno{(2)},$$ where
\$\forall n:\enskip\alpha_n\$ are complex
numbers, and
\$\forall n:\enskip\ket{n}_A\$ and
\$\ket{n}_B\$ for the same \$n\$ value
are referred to as {\bf
partners} in a pair of {\bf Schmidt
states}.\\

The term "Schmidt decomposition" can be
replaced by "Schmidt expansion"  or
"Schmidt form". To avoid confusion, we'll
stick to the first term throughout (as it
is usually done in the literature).\\

{\bf Theorem 3.} Expansion (1a) is a {\bf
Schmidt decomposition \IF} the
second-tensor-factor-space basis
\$\{\ket{n}_B:
\forall n\}\$ is an {\bf eigen-basis}
of the corresponding reduced density
operator
\$\rho_B\enskip\Big[\equiv\tr_A\Big(
\ket{\Psi}_{AB}\bra{\Psi}_{AB}\Big)\Big]\$:
$$\forall n:\quad\rho_B\ket{n}_B=
r_n\ket{n}_B,\quad 0\leq
r_n.\eqno{(3)}$$\\

{\bf Proof.} Let us evaluate
\$\overline{\bra{n}_A}\overline{\ket{n'}_A}\$
making use of (1b).
$$\overline{\bra{n}_A}\overline{\ket{n'}_A}=
\Big(\bra{\Psi}_{AB}\ket{n}_B\Big)
\Big(\bra{n'}_B\ket{\Psi}_{AB}\Big)=
\bra{\Psi}_{AB}\Big(\ket{n}_B
\bra{n'}_B\Big)\ket{\Psi}_{AB}=$$ $$
 \tr\Big((\ket{\Psi}_{AB}\bra{\Psi}_{AB})
(\ket{n}_B\bra{n'}_B)\Big)=
\tr_B\Big[\Big(\tr_A(\ket{\Psi}_{AB}\bra{\Psi}_{AB})\Big)
(\ket{n}_B\bra{n'}_B)\Big]=$$ $$
\tr_B\Big(\rho_B
(\ket{n}_B\bra{n'}_B)\Big)=\bra{n'}_B\rho_B\ket{n}_B.$$
The third equality in the above
derivation, where the expectation value
is rewritten as a suitable trace, is a
standard, textbook step. (Evaluating the
trace in a basis in which the relevant
state vector is one of the basis elements, the equality becomes obvious.) In the fourth equality the first partial-trace rule (cf Appendix B) was used.

We have obtained
 $$\overline{\bra{n}_A}\overline{\ket{n'}_A}=
\bra{n'}_B\rho_B\ket{n}_B.\eqno{(4)}$$

It is clear from relation (4) that the
vectors \$\{\overline{\ket{n}_A}:\forall
n\}\$ are orthogonal \IF \$\rho_B\$ is
diagonal, and this is the case \IF the
eigenvalue relations (3) are valid as
claimed.\hfill $\Box$\\

{\bf Corollary 2.} If one expands
\$\ket{\Psi}_{AB}\$ in a
second-subsystem basis like in (1a), then
the subsystem state (reduced density
operator) \$\rho_A\$ is given as a
mixture (1c). If, in addition, the
\$B$-subsystem basis is an eigen-basis of
\$\rho_B\$,  then (1c) is simultaneously
also a spectral decomposition of
\$\rho_A\$ (in terms of its eigen-vectors).\\

Now we turn to a special form of
Schmidt decomposition that is often
more useful. It is called canonical
Schmidt decomposition. It is due to the
fact that the non-trivial phase factors
of the non-zero coefficients \$\alpha_n\$
in
(2) can be absorbed either in the basis
elements in \$\cH_A\$ or in those in
\$\cH_B\$ (or partly in the former and
partly in the latter).\\

{\bf Definition 3.} If in a Schmidt
expansion (2) all \$\alpha_m\$ are {\bf
non-negative real
numbers}, then we write the expansion
in the following way:
$$\ket{\Psi}_{12}=\sum_ir_i^{1/2}
\ket{i}_A\ket{i}_B,\eqno{(5)}$$ and the
sum is confined to non-zero terms (one
is reminded of all this by the
replacement of the index \$n\$ by \$i\$
in this notation). Relation (5) is
called {\bf canonical Schmidt
decomposition}. (The term "canonical"
reminds of the form of (5), i. e., of
\$\forall i:\enskip r_i^{1/2}>0.\$)\\

Needless to say that every
\$\ket{\Psi}_{AB}\$ can be written in the
form of a canonical Schmidt
decomposition, and it is, of course,
non-unique.\\

{\bf Corollary 3.} Every canonical
Schmidt decomposition (5) is accompanied
by the {\bf spectral forms of the reduced
density operators}:
$$\rho_s=\sum_ir_i
\ket{i}_s\bra{i}_s,\quad s=A,B.
\eqno{(6a,b)}$$ (Note that the same
eigenvalues \$r_i\$ appear in (5) and in
the two spectral forms (6a) and (6b).
Note also that (6a) is the same as (1c)
if the RHS(1c) is determined by (1a), and \$\{\ket{n}_B:\forall n\}\$ is
an eigen-basis of \$\rho_B\$.)\\

{\bf Proof.} The Schmidt canonical
decomposition (5) allows the
straightforward evaluation $$
\rho_A\equiv\tr_B\Big(\ket{\Psi}_{AB}
\bra{\Psi}_{AB}\Big)=\sum_{i,i'}r_i^{1/2}
r_{i'}^{1/2}\tr_B\Big(\ket{i}_A\ket{i}_B\bra{i'}_A
\bra{i'}_B\Big)=$$
$$\sum_{i,i'}r_i^{1/2}
r_{i'}^{1/2}(\ket{i}_A\bra{i'}_A)\tr\Big(\ket{i}_B
\bra{i'}_B\Big)=RHS(6a)$$ (the first partial-trace rule was made use of). Relation (6b) is proved
symmetrically.\hfill $\Box$\\

One should note that the {\bf ranges}
\$\cR(\rho_s),\enskip s=A,B\$, of the
reduced
density operators \$\rho_s,\enskip
s=A,B\$ are {\bf equally dimensional}.
The common dimension is the number of
terms in a canonical Schmidt
decomposition (5). (It is sometimes
called the Schmidt rank of the given
bipartite state vector.)\\

We denote the {\bf range-projectors} of
the reduced density operators
\$\rho_s,\enskip s=A,B\$ by \$Q_s,\enskip
s=A,B\$. It is seen from (6a,b) that
$$Q_s=\sum_i\ket{i}_s\bra{i}_s,\quad
s=A,B
\eqno{(6c,d)}$$ are valid. The reduced
density
operators have {\it equal positive
eigenvalues}
\$\{ r_i>0:\forall i\}\$ (implying
equality of the multiplicities of the
distinct ones among them). The possible
zero eigenvalues may differ arbitrarily
(cf (6a,b)).\\

The Schmidt canonical decomposition was studied in \cite{FHMV'76}.\\

{\bf Corollary 4.} The following
relations are always valid:
$$\ket{\Psi}_{AB}=Q_s\ket{\Psi}_{AB},\quad
s=A,B.$$\\

{\bf Proof.} Since
\$Q_s=\sum_i\ket{i}_s\bra{i}_s,\enskip
s=A,B\$, the claim is obvious  when
\$\ket{\Psi}_{AB}\$ is written as a
canonical Schmidt decomposition
(5).\hfill $\Box$\\

{\bf Corollary 5.} One always has
$$\ket{\Psi}_{AB}\in\cR(Q_AQ_B).$$\\

{\bf Remark 1.} If we enumerate by \$j\$
the {\bf distinct} positive common
eigenvalues
\$\{r_j>0:\forall j\}\$ of
\$\rho_s,\enskip s=A,B\$, and by
\$Q_s^j,\enskip s=A,B\$
the corresponding {\bf eigen-projectors},
then one has the relations
$$\rho_s=\sum_jr_jQ_s^j,\quad s=A,B
,\eqno{(7a)}$$
$$\bar\cR(\rho_s)=\cR(Q_s)=\sum_j^{\oplus}\cR(Q_s^j)
\quad s=A,B.\eqno{(7b)}$$
$$\forall j:\quad
dim\Big(\cR(Q_A^j)\Big)=dim\Big(\cR(Q_B^j)\Big)<\infty
.\eqno{(7c)}$$ As to (7b), one should
note that \IF
\$dim\big(\cR(\rho_s)\Big)=\infty,\enskip
s=A,B\$, then the range \$\cR(\rho_s)\$
is a linear manifold that is not equal
but only dense in its topological closure
\$\bar\cR(\rho_s),\enskip s=A,B\$. The
symbol "$\oplus$" denotes orthogonal sum
of subspaces.

One should also note that all
positive-eigenvalue eigen-subspaces
\$\cR(Q_s^j)\$ are necessarily always
{\bf finite dimensional} ((7c)) because
\$\sum_ir_i=1\$ (a
consequence of the normalization of
\$\ket{\Psi}_{AB}\$), and hence no
positive-eigenvalue can have infinite
degeneracy. But there may be denumerably
infinitely many distinct positive
eigenvalues \$r_j\$.

We refer to (7a), (7b) and (7c) as the
{\bf subsystem picture} of
\$\ket{\Psi}_{AB}\$. It serves as a first layer of an underlying grounding for Schmidt decomposition.\\

{\bf Remark 2.} One can say that one has
a canonical Schmidt decomposition (5)
{\bf \IF} the expansion is bi-ortho-normal
and all expansion coefficients are
positive.\\

{\bf Remark 3.} A canonical Schmidt
decomposition (5) of any bipartite state
vector \$\ket{\Psi}_{AB}\$ is non-unique
because the eigen-sub-basis
\$\{\ket{i}_B:\forall i\}\$ of
\$\rho_B\$ spanning its range
\$\cR(\rho_B)\$
is {\bf non-unique}. Even if \$\rho_B\$
is
non-degenerate in all its positive
eigenvalues, there is the
non-uniqueness of the phase factors of
\$\ket{i}_B\$.\\

\section{Correlated Schmidt decomposition}

We investigate further the mentioned
non-uniqueness (see end of the preceding
section). In the canonical Schmidt
decomposition (5) it is clear that the
entanglement in \$\ket{\Psi}_{AB}\$ boils
down to the choice of the partner in the
terms of the decomposition.

We introduce explicitly this {\it choice
of a partner} keeping in mind the
subsystem picture (cf (7a)-(7c)). It
turns out that the best thing to do is to
define an {\bf antiunitary map} that
takes the
(topologically) closed range
\$\bar\cR(\rho_B)\$ onto the symmetrical
entity \$\bar\cR(\rho_A).\$

The map is called {\bf the
correlation operator}, and it is denoted
by
the symbol \$U_a\$ \cite{Varenna}, \cite{FHMV'76}, \cite{MVFHJMP}.\\

{\bf Definition 4.} If a canonical
Schmidt decomposition (5) is given, then
{\bf the two ON sub-bases} of equal
power \$\{ \ket{i}_B:\forall i\}\$ and
\$\{\ket{i}_A:\forall i\}\$ appearing in
it {\bf determine} an antiunitary, i.
e.,
antilinear and unitary, operator
\$U_a,\$ the {\bf correlation operator} -
a correlation entity inherent in the
given state vector \$\ket{\Psi}_{AB}$:
$$\forall i:\quad \ket{i}_A\equiv
\Big(U_a
\ket{i}_B\Big)_A. \eqno{(8)}$$\\

The correlation operator \$U_a,\$
mapping \$\bar\cR(\rho_B)\$ onto
\$\bar\cR(\rho_A),\$ is well defined by
(8) and by the additional requirements
of antilinearity (complex conjugation
of numbers, coefficients in any linear
combination on which the operator may
act) and  continuity (if the
bases are infinite). (Both these
requirements follow from that of
antiunitarity.) Preservation of every
scalar product up to complex
conjugation, which, by definition,
makes \$U_a\$ antiunitary, is easily
seen to follow from (8) and the
requirements of antilinearity and
continuity because \$U_a\$ takes a
basis into another one.\\

{\bf Definition 5.} On account of \D 4
and (8), any canonical Schmidt
decomposition (5) of any bipartite state
vector \$\ket{\Psi}_{AB}\$
can be written in the form
$$\ket{\Psi}_{AB}=
\sum_ir_i^{1/2}\Big(U_a\ket{i}_B\Big)_A\otimes
\ket{i}_B.\eqno{(9)}$$ This form is
called a {\bf correlated canonical
Schmidt decomposition}. (In
\cite{FHenvar1}, section 2, instead of the term 'correlated'  the term 'strong' was used.)

One should note that (9) contains all the
entities that appear in (5) plus
(explicitly) the correlation operator
\$U_a\$, which is implicitly contained in
(5). Expansion (9) makes explicit the
fact that the opposite-subsystem
eigen-sub-basis \$\{\ket{i}_A:\forall
i\}\$ in (5) is not just any such set of
vectors once the eigen-sub-basis
\$\{\ket{i}_B:\forall i\}\$ is chosen.\\

{\bf Theorem 4.} The correlation
operator \$U_a\$ is {\bf uniquely
determined} by the given (arbitrary)
bipartite state vector
\$\ket{\Psi}_{AB}$.\\

{\bf Proof.} Let
\$\{\ket{j,k_j}_B:\forall
k_j,\forall j\}\$ and
\$\{\ket{j,l_j}_B:\forall l_j,\forall
j\}\$ be two arbitrary eigen-sub-bases
of \$\rho_B\$ spanning
\$\bar\cR(\rho_B).\$ The vectors are
written with two indices, \$j\$
denoting the eigen-subspace
\$\cR(Q_B^j)\$, \$\forall j:\enskip
Q_B^j\equiv\sum_{k_j}\ket{j,k_j}_B
\bra{j,k_j}_B$, corresponding to the
eigenvalue \$r_j\$ of \$\rho_B\$ to which
the vector
belongs, and the other index \$k_j\$
(\$l_j\$) enumerates the vectors within
the eigen-subspace \$\cR(Q_B^j)\$ in case
the eigenvalue \$r_j\$ of \$\rho_B\$ is
degenerate, i. e., if its  multiplicity
is more than 1.

The proof goes as follows. Let
$$\forall j:\quad \ket{j,k_j}_B=
\sum_{l_j}U_{k_j,l_j}^{(j)}\ket{j,l_j}_B,$$
where \$\Big(U_{k_j,l_j}^{(j)}\Big)\$
are unitary sub-matrices. Then, keeping
\$U_a\$ one and the same, we can start
out with the correlated Schmidt
decomposition in the
\$k_j$-eigen-sub-basis, and after a few
simple steps (utilizing the
antilinearity of \$U_a\$ and the
unitarity of the transition
sub-matrices), we end up with the
correlated Schmidt decomposition (of the
same \$\ket{\Psi}_{AB}\$) in the
\$l_j$-eigen-sub-basis. Complex
conjugation is denoted by asterisk.

$$\ket{\Psi}_{AB}=\sum_jr_j^{1/2}\sum_{k_j}
\Big(U_a\ket{j,k_j}_B\Big)_A
\ket{j,k_j}_B=$$
$$\sum_jr_j^{1/2}\sum_{k_j}\Big\{
\Big(\sum_{l_j}\Big[\Big(U_{k_j,l_j}^{(j)}\Big)^*
\Big(U_a\ket{j,l_j}_B\Big)_A\Big]\otimes
\Big(\sum_{l_j'}
U_{k_j,l_j'}^{(j)}\ket{j,l_j'}_A\Big)_B\Big\}=$$
$$\sum_jr_j^{1/2}\sum_{l_j}\sum_{l_j'}\Big\{
\Big(\sum_{k_j}\Big(U_{k_j,l_j}^{(j)}\Big)^*U_{k_j,
l_j'}^{(j)}\Big)\Big(U_a\ket{j,l_j}_B\Big)_A
\otimes\ket{j,l_j'}_B\Big\}=$$
$$\sum_jr_j^{1/2}\sum_{l_j}\sum_{l_j'}
\Big\{
\delta_{l_j,l_j'}\Big(U_a\ket{j,l_j}_B\Big)_A
\otimes\ket{j,l_j'}_B\Big\}=
\sum_jr_j^{1/2}\sum_{l_j}\Big(U_a
\ket{j,l_j}_B\Big)_A\ket{j,l_j}_B.$$\hfill
$\Box$\\

It may seem that the uniqueness of
\$U_a\$ when
\$\ket{\Psi}_{AB}\$ is given is a
poor compensation for the trouble one
has treating an antilinear operator.
But the difficulty is more
psychological than practical, because
all that distinguishes an antiunitary
operator from a unitary one is

(i) its antilinearity - it
complex-conjugates
the numbers in any linear combination on
which it acts - and

(ii) its property that it
complex-conjugates every scalar product
(preserving its absolute value):
\$\bra{\psi}\ket{\phi}=
[(\bra{\psi}U_a^{\dag})(U_a\ket{\phi})]^*\$.
The full compensation comes, primarily from the insight in entanglement that \$U_a\$ furnishes, from its practical usefulness, and, at last but not at least, from its important physical meaning.\\

The physical meaning of the correlation operator \$U_a\$ is best discussed in the context of Schr\" odinger's steering (see three passages beneath relation (34) in subsection 6.3 below). One should realize that physical meaning in \QM comes always heavily packed in mathematics. One must discern the physics in the haze of the formalism. This is attempted below.\\

{\bf Remark 4.} If a correlated Schmidt
canonical expansion (9) is written down,
then it can be viewed in two opposite
ways:

(i) as a given bipartite state vector
\$\ket{\Psi}_{AB}\$ determining its two
inherent entities, the reduced density
operator \$\rho_B\$ in spectral form
(cf (6b)) and the correlation operator
\$U_a\$ (cf (8)), both relevant for
the entanglement in the state vector
(and one can read them in the given
expansion);

(ii) as a given pair \$(\rho_B,U_a)\$
(\$U_a\$ mapping antiunitarily
\$\bar\cR(\rho_B)\$ onto some equally
dimensional subspace of \$\cH_A\$)
determining a bipartite state vector
\$\ket{\Psi}_{AB}$.

The second view of correlated Schmidt
expansion allows a systematic
generation and classification of all
state vectors in \$\cH_A\otimes\cH_B\$
(cf \cite{constr.all.bip}).\\

{\bf Theorem 5.}  The expansion
coefficients
\$\{\overline{\ket{n}_A}:\forall n\}\$
in any subsystem-basis expansion (1a) can
be evaluated, besides by (1b), also
utilizing the reduced density operator
\$\rho_B\$ and the correlation operator
\$U_a\$ as follows: $$\forall n:\quad
\overline{\ket{n}_A}=\Big(U_a\rho_B^{1/2}\ket{n}_B\Big)_A.
\eqno{(10)}$$

{\bf Proof.} We substitute a canonical
Schmidt decomposition of
\$\ket{\Psi}_{AB}\$
in (1b) for an arbitrary \$n\$ value:
$$\overline{\ket{n}}_A=\bra{n}_B\ket{\Psi}_{AB}=
\bra{n}_B\Big(\sum_ir_i^{1/2}\ket{i}_A\ket{i}_B\Big)=
\sum_ir_i^{1/2}\bra{n}_B\ket{i}_B\times\ket{i}_A.
\eqno{(11a)}$$

On the other hand, evaluating the RHS of
(10) making use of the spectral form (6b)
of \$\rho_B\$ and of (8), we obtain:
$$U_a\rho_B^{1/2}\ket{n}_B=U_a\Big(\sum_ir_i^{1/2}
\ket{i}_B\bra{i}_B
\Big)\ket{n}_B=\sum_ir_i^{1/2}(\bra{i}_B\ket{n}_B)^*\times
(U_a\ket{i}_B)_A=$$
$$\sum_ir_i^{1/2}\bra{n}_B\ket{i}_B\times
\ket{i}_A.\eqno{(11b)}$$ The asterisk
denotes complex conjugation. It is
required by the antilinearity property of
the correlation operator.
 Comparing (11a) and (11b), we see that
 the RHS's are equal, hence so are the
 LHS's.\hfill $\Box$\\

Theorem 5, as it stands, is new with respect to previous work. Though, in \cite{FHMV'76} (relation (34) there) an analogous result was obtained, but the derivation was formulated and presented in the approach in which bipartite states are written as antilinear Hilbert-Schmidt mappings of \$\cH_B\$ into \$\cH_A\$. This approach is almost never used in the literature.\\

{\bf Remark 5.} Substituting (10) in
(1a) one obtains
 $$\ket{\Psi}_{AB}=\sum_n\Big(U_a\rho_B^{1/2}\ket{n}_B\Big)_A
 \otimes\ket{n}_B.\eqno{(12)}$$ This can
 be called a {\bf generalized correlated
 canonical Schmidt decomposition}. Note
 that the nearby subsystem basis
 \$\{\ket{n}_B:\forall n\}\$ is not
 necessarily an eigen-basis of
 \$\rho_B\$; it is arbitrary. This is how
 it is a generalization. Form (12) of
 expansion in a subsystem basis is relevant
 for Schr\"odinger's steering discussed
 in detail in subsection 6.3 below.\\

{\bf Remark 6.} Theorem 5 and relation (10) enables one to prove the uniqueness of the correlation operator \$U_a\$
 independently of \T 4. Namely, this
 uniqueness is a consequence of the
 uniqueness of the partial scalar product
 (proved in Appendix A).\\

{\bf Remark 7.} When a pair of ON
sub-bases
\$\{\ket{i}_B:\forall i\}\$ and
\$\{\ket{i}_A:\forall i\}\$ appearing in
a
canonical Schmidt decomposition (5)
is given, one can extend \$U_a\$ to the
entire \$\cH_B\$, denote the extended
operator as \$\bar U_a\$, and write
$$\bar U_a=
\sum_i\ket{i}_AK\bra{i}_B,\eqno{(13a)}$$
where \$K\$ is complex conjugation
(denoted by asterisk when acting on
numbers). \D (13a) is actually
symbolical. Its true meaning consists in
the following.
$$\forall \ket{\phi}_B\in\cH_B:\quad\bar
U_a\ket{\phi}_B=\Big(\sum_i\ket{i}_AK
\bra{i}_B\Big)\ket{\phi}_B\equiv
\sum_i(\bra{i}_B\ket{\phi}_B)
^*\ket{i}_A.\eqno{(13b)}$$ The extended
operator\$\bar U_a\$ acts as \$U_a\$ in
the range \$\cR(\rho_B)\$, and  it acts
as zero in the null space of \$\rho_B\$.
In other words, one can write $$\bar
U_a=U_aQ_B,\eqno{(13c)}$$ where \$Q_B\$
is the range projector of \$\rho_B\$.
Since  \$Q_B\$ projects onto the range,
it does not matter that \$U_a\$ is
defined only on the range.\\

{\bf Remark 8.} As one can easily see,
utilizing complete ON eigen-bases of
\$\rho_s\$, \$s=A,B\$ (cf (6a-d) and
(8)), one has
$$\rho_A=U_a\rho_BU_a^{-1}Q_A,
\quad \rho_B=U_a^{-1}\rho_AU_aQ_B.
\eqno{(14a,b)}$$ Thus, {\it the reduced density operators
are, essentially, "images" of each
other via the correlation operator}.
(The term "essentially" points to the
fact that the dimensions of the null
spaces are independent of each other.)
Property (14a,b) is called {\it twin
operators}. (More will be said about such
pairs of operators below, cf \D 6
below.)

In terms of {\it subspaces}, to (14a,b)
correspond the image-relations
$$\cR(Q_A)=U_a\cR(Q_B),\quad
\cR(Q_B)=U_a^{-1}\cR(Q_A).\eqno{(14c,d)}$$

One obtains an even more detailed view
when one takes into account the {\it
eigen-subspaces} \$\cR(Q_s^j)\$ of
\$\rho_s\$ corresponding to (the
common) distinct positive eigenvalues
\$r_j\$ of \$\rho_s,\$ where \$Q_s^j\$
projects onto the
\$r_j-$eigen-subspace, \$s=A,B\$ (cf the
subsystem picture (7a)-(7c)). Then
one obtains a view of
the {\bf entanglement} in a given
composite state \$\ket{\Psi}_{AB}\$ in
terms of the
so-called {\bf correlated subsystem
picture} \cite{Varenna}:
$$\rho_s=\sum_jr_jQ_s^j,\quad s=A,B
,\eqno{(15a,b)}$$ and in terms of
subspaces
$$\bar\cR(\rho_s)=\sum_j^{\oplus}\cR(Q_s
^j),\quad s=A,B,\eqno{(15c,d)}$$ where
\$"\oplus"\$ denotes an orthogonal sum
of subspaces.

Further, as it is also straightforward to
see in eigen-bases of \$\rho_s,\enskip
s=A,B\$,
$$\forall j:\quad
\cR(Q_A^j)=U_a\cR(Q_B^j),\quad
\cR(Q_B^j)=U_a^{-1}\cR(Q_A^j).
\eqno{(15e,f)}$$

In words, the correlation operator
makes not only the ranges of the
reduced density operators "images" of
each other, but also all
positive-eigenvalue eigen-subspaces of
the reduced density operators. In
other words, the correlation operator
\$U_a$, making the reduced density
operators \$\rho_s,\enskip s=A,B\$
'images' of each other, makes also the
eigen-decompositions of the
ranges \$\cR(\rho_s),\enskip s=A,B\$
'images' of each other.

The relations (14a)-(14d) and (15a)-(15f)
constitute the {\bf correlated subsystem
picture} of the given state vector
\$\ket{\Psi}_{AB}\$ in terms of operators
and corresponding subspace state
entities. This is the second layer in the underlying grounding of the (correlated) Schmidt decomposition.\\

\section{Twin-correlated Schmidt decomposition}

In the correlated subsystem picture of a
given bipartite state vector
\$\ket{\Psi}_{AB}\$ (in the preceding section) we have searched
for a comprehension of entanglement and
its canonical form, but doing so we have investigated {\it only state
entities} \$\rho_A,\rho_B,U_a\$ . Now, we
introduce observables that can contribute
to the theory by enriching and broadening
our understanding.\\

{\bf Lemma 1.} Let \$\ket{\Psi}_{AB}\$ be
a bipartite state vector, \$\rho_A\$ its
first-subsystem reduced density operator,
\$Q_A\$ the range projector of the
latter, and \$O_A=\sum_ko_kP_A^k\$ a
first-subsystem observable in
spectral form. Let further \$P_A^{\not=
0}\$ be the sum of all those
eigen-projectors \$P_A^k\$ of \$O_A\$
that do not nullify \$\ket{\Psi}_{AB}\$.
Then, \$P_A^{\not= 0}Q_A=Q_A\$, i. e.,
\$P_A^{\not= 0}\geq Q_A\$, or, in words,
\$P_A^{\not= 0}\$ is 'larger' than \$Q_A\$
or equivalently \$\cR(P_A^{\not=
0})\supseteq\cR(Q_A)\$.\\

{\bf Proof.} One can write
$$\rho_A\equiv\tr_B\Big(\ket{\Psi}_{AB}\bra{\Psi}_{AB}\Big)=
\tr_B\Big((\sum_kP_A^k)\ket{\Psi}_{AB}
\bra{\Psi}_{AB}\Big)=P_A^{\not=
0}\rho_A$$ (the second partial-trace rule in Appendix B has been utilized).

Taking an eigen-sub-basis
\$\{\ket{i}_A:\forall i\}\$ of \$\rho_A\$
spanning its range, one can
further write \$\rho_A\$ in spectral form
and one obtains
$$\sum_ir_i\ket{i}_A\bra{i}_A=\sum_ir_i
P_A^{\not= 0}\ket{i}_A\bra{i}_A,\quad\forall i:\enskip r_i>0.$$
Applying this to an eigen-vector
\$\ket{\bar i}_A\$ corresponding to
\$r_{\bar i}>0\$, one obtains
\$r_{\bar i}\ket{\bar i}_A=r_{\bar i}P_A^{\not=
0}\ket{\bar i}_A\$. Finally, since
\$Q_A=\sum_i\ket{i}_A\bra{i}_A\$, the
claimed relation follows.\hfill $\Box$\\

{\bf Definition 6.} Let
\$O_A\equiv\sum_ka_k P_A^k\$ and \$O_B\equiv\sum_lb_l
P_B^l\$ be opposite-subsystem
Hermitian operators (observables)
in spectral form. If one
can renumerate all eigen-projectors
\$P_A^k\$ and \$P_B^l\$ that do not
nullify the given composite state vector
\$\ket{\Psi}_{AB}\$ by a common index, e. g. \$m\$, so that $$\forall m:\quad
P_A^m\ket{\Psi}_{AB}=P_B^m
\ket{\Psi}_{AB}\eqno{(16)}$$ is valid,
then the operators \$O_a\$ and \$O_B\$
are said to be {\bf twin operators} or
{\bf twin observables} in
\$\ket{\Psi}_{AB}\$.Twin projectors will
also be called twin events.\\

In \cite{twinsPR} twin observables were
called 'physical twins', and also
'algebraic twins; were mentioned. They
were defined by
\$O_A\ket{\Psi}_{AB}=O_B\ket{\Psi}_{AB}\$.\\

{\bf Remark 9.} Introducing \$P_s^{\not=
0}\equiv\sum_mP_s^m,\enskip s=A,B\$, Lemma
1 implies \$P_s^{\not= 0}Q_s=Q_s\$, i. e.,
that \$Q_s\$ is a sub-projector of
\$P_s^{\not= 0}:\enskip Q_s\leq P_s^{\not=
0}\$, or equivalently,
\$\cR(Q_s)\subseteq\cR(P_s^{\not=
0}),\enskip s=A,B.\$ Further, we can
define \$P_s^{=0},\enskip s=A,B\$ as the
sum of all nullifying eigen-projectors:
\$P_A^{=0}\equiv\sum_{k'}P_A^{k'}\$, where \$\forall k':\enskip
P_A^{k'}\ket{\Psi}_{A,B}=0\$, and
symmetrically for subsystem B. Then it
further follows that \$\forall k':\enskip P_s^{k'}\leq P_s^{=0}\leq
Q_s^c\$, where \$Q_s^c\equiv I_s-Q_s\$ is
the null-projector of \$\rho_s,\enskip
s=A,B\$.\\

{\bf Proposition 1.} The corresponding
results \$a_m\$ and \$b_m\$ of subsystem
\m s of twin observables are {\bf equally
probable} and {\bf ideal \M } causes {\bf
equal change} of the bipartite state:
$$\forall
m:\quad\bra{\Psi}_{AB}P_A^m\ket{\Psi}_{AB}=
\bra{\Psi}_{AB}P_B^m\ket{\Psi}_{AB},\eqno{(17)}$$
$$\ket{\Psi}_{AB}\bra{\Psi}_{AB}\quad
\rightarrow\quad\sum_m\Big(P_A^m\ket{\Psi}_{AB}
\bra{\Psi}_{AB}P_A^m\Big)=\sum_m
\Big(P_B^m\ket{\Psi}_{AB}\bra{\Psi}_{AB}
P_B^m\Big).\eqno{(18)}$$\\

{\bf Proof} follows obviously from \D 6
and relation (16).\hfill $\Box$\\

{\bf Theorem 6.} If \$O_A\$ and \$O_B\$
are twin operators (cf \D 6), then each
of their non-nullifying eigen-projectors
\$P_s^m,\enskip s=A,B\$ {\bf commutes}
with the corresponding reduced density
operator $$\forall m:\quad
[P_A^m,\rho_A]=0.\qquad
[P_B^m,\rho_B]=0.\eqno{(19a,b)}$$\\

{\bf Proof. } Straightforward evaluation,
utilizing (16) and both partial-trace rules from Appendix B,
gives:$$P_A^m\rho_A=P_A^m\tr_B\Big(\ket{\Psi}_{AB}
\bra{\Psi}_{AB}\Big)=\tr_B\Big((P_A^m\ket{\Psi}_{AB})
\bra{\Psi}_{AB}\Big)=$$
$$\tr_B\Big((P_B^m\ket{\Psi}_{AB})
\bra{\Psi}_{AB}\Big)=\tr_B\Big(\ket{\Psi}_{AB})
(\bra{\Psi}_{AB}P_B^m)\Big)=$$ $$
\tr_B\Big(\ket{\Psi}_{AB})
(\bra{\Psi}_{AB}P_A^m)\Big)=\Big[
\tr_B\Big(\ket{\Psi}_{AB})
(\bra{\Psi}_{AB}\Big)\Big]P_A^m=\rho_AP_A^m$$
The symmetrical claim is
proved symmetrically.\hfill $\Box$\\

We now state and prove (for the reader's
convenience) a basic claim of \QM that is
crucial for our further development of
the correlated subsystem picture
(elaborated in the preceding section).

{\bf Lemma 2.} Let \$O=\sum_ko_kP_k\$ and
\$\bar O=\sum_l\bar o_l\bar P_l\$ be two
{\it commuting} hermitian operators (each
with a purely discrete spectrum) in spectral form. Then also
$$\forall k,\enskip\forall l:\quad
[P_k,\bar P_l]=0.$$

{\bf Proof.} Let \$\ket{k,q_k}\$ be a
complete ON eigen-basis of
\$O\$: \$\forall k,q_k:\enskip
O\ket{k,q_k}=o_k\ket{k,q_k}\$. Then
\$O(\bar O\ket{k,q_k})=\bar
OO\ket{k,q_k}=
o_k(\bar O\ket{k,q_k})\$. Hence,
\$P_k(\bar O\ket{k,q_k})=(\bar
O\ket{k,q_k})=\bar OP_k\ket{k,q_k}\$.
Further, for \$k'\not= k\$, \$P_k(\bar
O\ket{k',q_{k'}})=
P_kP_{k'}(\bar O\ket{k',q_{k'}})=0=
\bar OP_k\ket{k',q_{k'}}\$. Thus,
\$\forall k:\enskip [P_k,\bar O]=0\$.
Applying this result to the last commutation itself, one finally obtains \$\forall k,l:\enskip
[P_k,\bar P_l]=0\$ as claimed.\hfill
$\Box$\\

{\bf Definition 7.} Let \$O_B=\sum_ka_kP_B^k\$ be a nearby-subsystem
observable that {\bf commutes} with the corresponding reduced density operator \$\rho_B\$ of
a given bipartite state vector \$\ket{\Psi}_{AB}\$. We re index the non-nullifying eigen-projectors of \$O_B\$ by \$m\$. Then, according to \L 2, each eigen-projector \$P_B^m\$ of \$O_B\$ commutes with \$Q_B^c\$, the null-projector of \$\rho_B\$ (cf \R 9), because it is also the eigen-projector of \$\rho_B\$ corresponding to its zero eigenvalue. This implies that it also commutes with \$Q_B\$ because the latter is ortho-complementary to \$Q_B^c\$. Hence, for each value of \$m\$, we can define the {\bf minimal sub-projector} \$P_B^{min,m}\$ that acts on \$\ket{\Psi}_{AB}\$ equally as \$P_B^m\$. Equivalently: $$\forall m:\quad P_B^{min,m}\ket{\Psi}_{AB}=P_B^m \ket{\Psi}_{AB},\qquad P_B^{min,m}\leq P_B^m,\enskip P_B^{min,m}\leq Q_B.\eqno{ (20a)}$$ Naturally, $$\forall m:\quad P_B^{min,m}=P_B^mQ_B=Q_BP_B^mQ_B.\eqno{ (20b)}$$ Finally, we can define $$O_B^{min}\equiv\sum_ma_mP_B^{min.m}\eqno{ (20c)}$$ and call it the {\bf minimal part} of \$O_B\$.\\

{\bf Proposition 2.} If \$O_B=\sum_ka_kP_B^k\$ commutes with \$\rho_B\$, then the corresponding minimal
operator \$O_B^{min}\$ can be obtained
as follows: $$O_B^{min}\equiv O_BQ_B. \eqno{(20d)}$$

{\bf Proof.} We write
\$O_B=(\sum_m a_mP_B^m)+\sum_{k'}a_{k'}P_B^{k'}\$. Here by \$P_B^{k'}\$ are denoted the nullifying eigen-projectors of \$O_B\$ (cf \R 9). Then (20b), (20c), and \R 9 imply $$O_BQ_B=\Big(\sum_m
a_mP_B^m+\sum_{k'}a_{k'}P_B^{k'}\Big)Q_B=$$ $$\Big(\sum_m
a_mP_B^{min,m}+\sum_{k'}a_{k'}P_B^{k'}Q_B^c\Big)Q_B=
\sum_ma_mP_B^{min.m}=O_B^{min}.$$\hfill $\Box$\\

{\bf Remark 10.} Commutation (19b) and \R 9, which claims that \$Q_B=Q_B\sum_mP_B^m\$, and since \$\forall j:\enskip Q_B^jQ_B=Q_B^j\$, the former relation implies \$\forall j:\enskip Q_B^j\sum_mP_B^m=Q_B^j\$, in conjunction with (20b), lead to the following spectral operator
decomposition:
$$\rho_B=\sum_jr_j\sum_mQ_B^jP_B^{min,m},\eqno{(21a)}$$
or in terms of the corresponding
subspaces
$$\cR(Q_B)=\sum_j^{\oplus}\sum_m^{\oplus}
\Big(\cR(Q_B^j)\cap\cR(P_B^{min,m}\Big).\eqno{(21b)}$$
Naturally, the RHS of (21a) may contain
zero operator terms, and on the RHS of
(21b) may appear corresponding zero
subspaces.\\

{\bf Remark 11.} As it is well known, the
commutation relations (19b) and (19a)
imply that there exist common eigen-bases
of \$\rho_B\$ and \$O_B^{min}\$ in
\$\cR(Q_B)\$ as well as of \$\rho_A\$ and
\$O_A^{min}\$ in \$\cR(Q_A)\$. We are
primarily
interested in the former. Let by
\$(jm)'\$ be denoted a pair of indices for which
\$Q_B^jP_B^{min,m}\not= 0\$. We introduce a third index
\$q_{(jm)'}\$ to enumerate the
ortho-normal vectors in the corresponding non-zero
subspaces
\$\cR(Q_B^j)\cap\cR(P_B^{min,m})\$.\\

{\bf Remark 12.} The decomposition without zero terms is
$$Q_B=\sum_{(jm)'}Q_B^jP_B^{min,m}=
\sum_{(jm)'}\sum_{q_{(jm)'}}
\ket{(jm)'q_{(jm)'}}_B\bra{(jm)'q_{(jm)'}}_B.\eqno{(22)}$$\\

{\bf Definition 8.} Expanding a given
bipartite state \$\ket{\Psi}_{AB}\$ in
the subsystem sub-basis appearing in
(22), we obtain, what we call, the {\bf
twin-correlated canonical Schmidt
decomposition}:
$$\ket{\Psi}_{AB}=\sum_{(jm)'}\sum_{q_{(jm)'}}
r_j^{1/2}\ket{(jm)'q_{(jm)'}}_A\ket{(jm)'q_{(jm)'}}_B,\eqno{(23a)}$$
with $$\forall (jm)'q_{(jm)'}:\quad
\ket{(jm)'q_{(jm)'}}_A=
\Big(U_a\ket{(jm)'q_{(jm)'}}_B\Big)_A\eqno{(23b)}$$
(cf the correlated canonical Schmidt
decomposition (5)). If the role of the
correlation operator \$U_a\$ is not made
explicit in (23a) or, equivalently, if (23b) is not
joined to it, i. e., (23a) itself (as it
stands) we call {\bf twin-adapted
canonical Schmidt decomposition}.\\

As a consequence of (23b), one has
$$\forall (jm)':\quad
Q_A^jP_A^{min,m}=U_a\Big(Q_B^jP_B^{min,m}\Big)U_a^{-1}Q_A,
\eqno{(24a)}$$
$$\forall (jm)':\quad
Q_B^jP_B^{min,m}=U_a^{-1}\Big(Q_A^jP_A^{min,m}\Big)
U_aQ_B.
\eqno{(24b)}$$\\

The following result is another obvious
consequence of (23b).\\

{\bf Theorem 7.} If \$O_s^{min},\enskip
s=A,B\$ are minimal twin observables for
\$\ket{\Psi}_{AB}\$, then
$$\forall m:\quad
P_B^{min,m}=\sum_j'\sum_{q_{(jm)'}}\ket{(jm)'q_{(jm)'}}_B
\bra{(jm)'q_{(jm)'}}_B,\eqno{(25a)}$$
$$\forall m:\quad
P_A^{min,m}=\sum_j'\sum_{q_{(jm)'}}\ket{(jm)'q_{(jm)'}}_A
\bra{(jm)'q_{(jm)'}}_A,\eqno{(25b)}$$ where the prim the sum over \$j\$ denotes restriction to those terms in which \$j\$ with the given \$m\$ gives a non-zero subspace in (21b). Further,
$$\forall m:\quad
P_A^{min,m}=U_aP_B^{min,m}U_a^{-1}Q_A,\eqno{(26a)}$$
$$\forall m:\quad
P_B^{min,m}=U_a^{-1}P_A^{min,m}U_aQ_B.\eqno{(26b)}$$\\

Relations (22), (23b), (24a,b) and
(26a,b) constitute the {\bf
twin-correlated subsystem picture}. It is the third and most intricate layer of the underlying foundation of Schmidt  decomposition. It completes the {\it correlated subsystem picture}
(see (14a)-(14d) and (15a)-(15f))
by the pair of minimal twin
observables \$O_A^{min},\enskip
O_B^{min}\$, and the latter picture was,
in turn, a completion of the {\it subsystem picture} (cf
(7a)-(7c)) by the correlation
operator.\\

The original articles \cite{JMPstructure} - \cite{twinsPR}, which have been reviewed here, did not present the third layer of foundation sufficiently precisely and transparently. Therefore, a completely new and different derivation is given in this section.\\

One may wonder if there may exist two
different observables \$O_A\$ and \$\bar
O_A\$ both twins with one and the same
opposite-subsystem observable \$O_B\$ in
a given \$\ket{\Psi}_{AB}\$.\\

{\bf Proposition 3.} If \$O_A\$ and
\$\bar O_A\$ are both twin observables
with one and the same opposite-subsystem
observable \$O_B\$, then
$$O_A^{min}=\bar O_A^{min}.$$

{\bf Proof} follows immediately from
(26a).\hfill $\Box$\\

{\bf Remark 13.} Thus, in this case, one can have
\$O_A\not=\bar O_A\$ only if \$\rho_A\$
is singular, and then the only difference
is in the terms \$P_A^mQ_A^c\$, where
\$Q_A^c\equiv I_A-Q_A\$ is the null-space projector of \$\rho_A\$. The
operators \$P_A^mQ_A^c\$ are
sub-projectors of  \$Q_A^c\$. These terms in the projectors
\$P_A^m=P_A^mQ_A+P_A^mQ_A^c\$ nullify
$\ket{\Psi}_{AB}\$. Taking \$O_A\$ or
\$\bar O_A\$ means no difference for the
entanglement in \$\ket{\Psi}_{AB}\$
because the latter takes place between
\$\cR(Q_B)\$ and \$\cR(Q_A)\$ (with no
regard to the null spaces of
\$\rho_s,\enskip s=A,B\$).\\

The minimal form of a discrete subsystem Hermitian operator that commutes with the corresponding reduced density operator of the given bipartite state vector \$\ket{\Psi}_{AB}\$ (cf \D 7 and \P 2) was not defined explicitly in previous work. Hence, the presentation there of this last and most intricate form of  Schmidt decomposition and its underlying entanglement foundation was not so transparent. In the present exposition there is new insight and there are new results.\\

One may wonder which observables \$O_B\$ do have a twin observable in the given bipartite state.

{\bf Theorem 8.} Let \$\ket{\Psi}_{AB}\$
be any bipartite state vector and let
\$O_B\equiv\sum_lb_l P_B^l\$ be an observable for the nearby
subsystem B. It has a twin observable \$O_A\$ \IF

{\bf A)} It, as an operator,
{\bf commutes} with the corresponding
reduced density operator: \$\rho_B\equiv
\tr_A\Big(\ket{\Psi}_{AB}\bra{\Psi}_{AB}\Big)\$,
\$[O_B,\rho_B]=0\$. Then {\bf there
exists a unique minimal twin observable}
\$O_A^{min}\$.

{\bf B)} If the bipartite state is expanded in an eigen-basis \$\{\ket{l,q_l}_B:\forall l,q_l\}\$ of \$O_B\$ $$\ket{\Psi}_{AB}=\sum_l\sum_{q_l}
\overline{\ket{l,q_l}}_A\ket{l,q_l}_B$$ the
'expansion coefficients' satisfy the orthogonality conditions: \$\overline{\bra{l,q_l}}_A\overline{\ket{l',q_{l'}}}_A=0\$
whenever \$l\not= l'\$.\\

{\bf Proof A)} follows in a straightforward
way from (22), for which the commutation
of \$O_B\$ with \$\rho_b\$ is sufficient
(cf Lemma 2). Then, with the help of
(23b), the eigen-projectors
\$(P_A^{min})^m\$ are defined by
(25b).

{\bf B)} Obvious. hfill $\Box$\\

One may further wonder if it can happen that
\$[O_B,\rho_B]=0\$, one expands
\$\ket{\Psi}_{AB}\$ in the common
eigen-basis of these two operators and one
does not obtain a twin-adapted Schmidt
decomposition of the bipartite state.\\

{\bf Remark 14.} The answer is NO: it
cannot happen. One necessarily obtains a
twin-adapted Schmidt decomposition in
terms of \$O_B^{min}\$ and
\$O_A^{min}\equiv\sum_mo_m\Big(U_aP_B^{min,m}U_a^{-1}
\Big)Q_A\$ (cf \D 7 and \P 2), where
\$Q_A=\sum_i\Big(U_a\ket{i}_B\Big)_A
\Big(\bra{i}_BU_a^{\dag}\Big)_A\$, and
the eigenvalues \$\{o_m:\forall m\}\$ are
arbitrary distinct non-zero real numbers
(they are irrelevant).\\

One may also wonder if there exits a
bipartite state that has no twin
observables. The answer is again: NO.
Formally, the reduced density operators
\$\rho_s,\enskip s=A,B\$ themselves are
twin operators, as obvious in the
canonical Schmidt decomposition (cf (5)
). They, or any other Hermitian operators
with the same eigen-projectors, can be
viewed as minimal (in the sense of \D 7) twin observables.\\


\section{Distant measurement and EPR states}

The 'correlation operator as an
entanglement entity' approach furnished a
specific view of a historically important
notion: the EPR paradox.\\


\subsection{Distant \M }

Let any bipartite state vector
\$\ket{\Psi}_{AB}\$ be given, and let
\$O_A=\sum_ma_mP_A^m+O'_A\$ and
\$O_B=\sum_mb_mP_B^m+O'_B\$ be twin
observables in it (cf \D 6). The relations \$O'_A\ket{\Psi}_{AB}=0=O'_B\ket{\Psi}_{AB}\$ are valid.

The change of state in non-selective
\cite{Kaempffer} (when no definite-result
sub-ensemble is selected)  ideal \M
\cite{Lud}, \cite{Messiah}, \cite{Laloe}
  $$\ket{\Psi}_{AB}\bra{\Psi}_{AB}\quad
\rightarrow\quad\sum_m
\Big(P_B^m\ket{\Psi}_{AB}\bra{\Psi}_{AB}
P_B^m\Big)\eqno{(27)}$$ can be caused, in
principle,  by {\bf direct \m } on the
nearby subsystem \$B\$. Further, this
composite-system change of state implies
the ideal-\M change of state
$$\rho_B\quad\rightarrow\quad\sum_m\Big(P_B^m
\rho_BP_B^m\Big)$$ on the nearby
subsystem B (obtained when the partial
trace over subsystem A is taken).

In this case, by the very definition of
subsystem \m , there is {\bf no
interaction} between the measuring
instrument and the distant subsystem
A.\\

{\bf Proposition 4.} In spite of lack of
interaction with the distant subsystem A
in the composite-system change-of-state
(27), this subsystem  nevertheless
undergoes the ideal-\M change
$$\rho_A\quad\rightarrow\quad\sum_m\Big(P_A^m
\rho_AP_A^m\Big)\eqno{(28)}$$ due to the
{\bf entanglement} in \$\ket{\Psi}_{AB}\$.\\

{\bf Proof.} The change is
implied by (18), and seen by taking the
partial trace over subsystem B.\hfill $\Box$\\

{\bf Definition 9.} Change
(28) is said to be due to {\bf distant \M } (on the distant subsystem A)
\cite{FHMV'76}.\\

{\bf Remark 15.} It has been proved in
\cite{SubsystMeas} that the ideal change
(28) on the distant subsystem A
can be caused by {\bf any exact subsystem
\M } of the {\bf twin observable} on the
nearby subsystem B. The entanglement in
\$\ket{\Psi}_{AB}\$ does not distinguish,
as far as influencing the distant
subsystem is concerned, ideal \m ,
non-ideal nondemolition (synonyms:
predictive, first-kind, repeatable) \M
and even demolition (synonyms:
retrodictive, second-kind,
non-repeatable) \m s on the nearby
susbsystem as long as they are exact \m s.\\

{\bf Remark 16.} One should notice that
{\bf distant \M is always ideal \m .}
Moreover, the {\bf non-selective version
does not change the state} of the
opposite distant subsystem A at all.
Namely, on account of the commutation
\$\forall m:\enskip
[P_A^m,\rho_A]=0\$ (cf
(19a) in \T 6), one has
$$\sum_mP_A^m\rho_AP_A^m=\sum_m\rho_AP_A^m=
\rho_A\sum_mP_A^m=\rho_A(\sum_mP_A^m+\sum_{\bar
k}P_A^{\bar k})=\rho_A,$$ (cf Remark 9).
Hence, {\bf only the selective version of
distant \M may change the distant
state}.\\

{\bf Remark 17.} One may further write
$$\rho_A=\sum_mP_A^m\rho_AP_A^m=
\sum_m[\tr(\rho_BP_B^m)]\times\Big(
P_A^m\rho_AP_A^m\Big/[\tr(\rho_AP_A^m)]\Big)$$
and view mathematically \$\rho_A\$ as an
{\bf orthogonal mixture of substates}
(selected subensembles empirically) each
predicting a definite value of \$O_A\$.
The {\bf selective distant \m s} reduce
\$\rho_A\$ to the corresponding
state term. Since non-selective \M is actually the entirety of all selective \m s, the true physical meaning of the change (28) is in making the term states available to selective \m .\\

{\bf Remark 18.} Let
\$\rho_A=\sum_mw_m\rho_A^m\$ (\$w_m\$ being statistical weights: \$\forall m:\enskip w_m\geq 0,\enskip\sum_mw_m=1\$) be an {\it arbitrary orthogonal} decomposition
of the distant state \$\rho_A\$. It can be realized
by non-selective distant \M caused by a
suitable subsystem \M on the nearby
subsystem B. Namely, the range projectors
\$Q_A^m\$ of the term states \$\rho_A^m\$ are orthogonal. Defining \$\forall
m:\enskip P_A^{min,m}\equiv Q_A^m\$ and
\$O_A\equiv\sum_ma_mP_A^{min,m}\$ (\$a_m\$ any distinct real numbers), one
has the commutation \$[O_A,\rho_A]=0\$,
and, according to \T 8 (reading it in
reverse), there exists a minimal twin
observable \$O_B\$ for the opposite
subsystem. Its \M gives rise to the
distant \M of \$O_A\$, and hereby to the
orthogonal state decomposition that we
have started with.\\

Let us for the moment forget about twin observables, and consider more general ones.\\

{\bf Remark 19.} {\it Non-selective \M }
of any nearby-subsystem observable
\$O_B=\sum_lb_lP_B^l\$ gives rise to a
distant state decomposition
$$\rho_A\equiv\tr_B\Big(\ket{\Psi}_{AB}
\bra{\Psi}_{AB}\Big)=\sum_l\tr_B\Big(P_B^l(\ket{\Psi}_{AB}
\bra{\Psi}_{AB})\Big)=$$
$$\sum_l\tr_B\Big(P_B^l(\ket{\Psi}_{AB}
\bra{\Psi}_{AB})P_B^l\Big)=$$
$$\sum_l\bra{\Psi}_{AB}P_B^l
\ket{\Psi}_{AB}\times
\tr_B\Big(P_B^l(\ket{\Psi}_{AB}
\bra{\Psi}_{AB})P_B^l\Big)\Big/\Big[\tr\Big(P_B^l(\ket{\Psi}_{AB}
\bra{\Psi}_{AB})P_B^l\Big)\Big].$$
(Idempotency and the first partial-trace
rule - cf Appendix B - have been made use of.) Note that {\it selective \M } of the
same nearby subsystem observable gives,
by, what is called, distant preparation, a term state in the
above distant state decomposition. The latter itself is a way of writing \$\rho_A\$ as a mixture.\\

{\bf Remark 20.} Clearly, a subsystem
\M of a twin observable
\$O_B=\sum_lb_lP_B^l\$ in a given state
\$\ket{\Psi}_{A,B}\$ (cf \D 6) measures actually the corresponding minimal observable
\$O_B^{min}=\sum_mb_mP_B^{min,m}\$ (cf \D 7 and \P 2).
But, on account of the correlation
operator as an entanglement entity
contained in the bipartite state,
simultaneously and {\it ipso facto} also
the distant twin observable
\$O_A^{min}=\sum_ma_mP_A^{min,m}=
\sum_ma_m
\Big(U_aP_B^{min,m}U_a^{-1}\Big)\$ is
distantly measured. This makes the role of entanglement transparent.\\

To my knowledge it is an open question if the counterpart of \R 18 holds true for non-orthogonal decompositions of \$\rho_A\$, i. e., if every such decomposition can be given rise to by \M of some nearby-subsystem observable.\\


\subsection{EPR states}

{\bf Definition 10.} If a bipartite state
vector \$\ket{\Psi}_{AB}\$ allows distant
\M of two mutually incompoatible
observables (non-commuting operators)
\$O_A\$ and \$\bar O_A\$, then we say
that we are dealing with an {\bf EPR
state} (following the seminal
Einstein-Podolsky-Rosen article
\cite{EPR}).\\

{\bf Theorem 9.} A state
\$\ket{\Psi}_{AB}\$ is an EPR one \IF at
least one of the positive eigenvalues
\$r_j\$ of \$\rho_B\enskip
\Big(\equiv\tr_A\ket{\Psi}_{AB}\bra{\Psi}_{AB}\Big)\$
is degenerate, i. e., has multiplicity at
least two. This amounts to some
repetition in the expansion coefficients
\$r_i^{1/2}\$ in the canonical Schmidt
decomposition (5).\\

{\bf Proof.} Considering the
twin-correlated subsystem picture (cf
(22), (23b), (24a,b), and (26a,b)), it is straightforward to see that if at least one non-zero
subspace
\$\cR\Big(Q_B^jP_B^{min,m}\Big)\$, indexed by
\$(jm)'\$, is two or more dimensional,
then, and only then, one can have two
different eigen-bases
\$\{\ket{(jm)'q_{(jm)'}}_B:\forall
(jm)',\forall q_{(jm)'}\}\$ and
\$\{\overline{\ket{(jm)'r_{(jm)'}}_B}:\forall
(jm)',\forall r_{(jm)'}\}\$ so that the
correlation operator \$U_a\$ can
determine the corresponding (also
different) eigenbases
\$\{\Big(U_a\ket{(jm)'q_{(jm)'}}_B\Big)_A:\forall
(jm)',\forall q_{(jm)'}\}\$  and
\$\{\Big(U_a\overline{\ket{(jm)'r_{(jm)'}}_B}\Big)_A:\forall
(jm)',\forall r_{(jm)'}\}\$ of distant
incompatible minimal observables
\$O_A^{min}\$ and \$\bar
O_A^{min}\$.\hfill $\Box$\\

The original EPR paper \cite{EPR}
discussed the two-particle state
\$\ket{\Psi}_{AB}\$ defined by a fixed
value \$\vec P\$ of the total linear
momentum \$\hat{\vec p_A}+\hat{\vec
p_B}=\vec P\$, where \$\hat{\vec
p_s},\enskip s=A,B\$ are the  particle
linear momentum vector operators, and a
fixed value \$\vec r\$ of the relative
radius vector \$\hat{\vec r_A}-\hat{\vec
r_B}=\vec r\$. (For clarity, this time operators are denoted with hats to distinguish them from fixed values of vectors.)

The discussion went essentially as
follows: If one performs a position \M of
the nearby particle B and obtains the
value \$\vec r_B\$, then {\it ipso facto}
the distant particle A acquires without
interaction, via distant \m , the value
\$\vec r_A\equiv\vec r+\vec r_B\$. On the
other hand, as an alternative, one can
perform a linear momentum \M of the
nearby particle with a result \$\vec
p_B\$ and also obtain, by distant \m , a
definite value of the linear momentum
\$\vec p_A=\vec P-\vec p_B\$ of the
distant particle without interaction.

The authors found this conclusion
paradoxical in view of the contention
that \QM was complete, and
\$\ket{\Psi}_{AB}\$ did not contain the
mentioned values obtained without
interaction (with a 'spooky' action as
Einstein liked to say), and, moreover, it could not contain the two
incompatible values simultaneously as
valid for one and the same pair of
particles because position and linear momentum are incompatible.

As a slight formal objection, one may notice that the mentioned fixed values
of the total linear momentum and the
relative radius vector belong to
continuous spectra, and the corresponding
state is of infinite norm (a generalized
vector). Bohm pointed out
\cite{Bohm} that one can easily escape
this formal difficulty by taking for
\$\ket{\Psi}_{AB}\$ not the original EPR
state described above, but the well known
singlet two-particle spin state
$$\ket{\Psi}_{AB}\equiv (1/2)^{1/2}\Big(
(\ket{+}_A\ket{-}_B-\ket{-}_A\ket{+}_B)\Big),
\eqno{(29)}$$ where \$+\$ and \$-\$
denote spin-up and spin-down respectively
along any axis. For the same
\$\ket{\Psi}_{AB}\$ given by (29) one can
choose either the z-axis or the x-axis,
and make an argument in complete analogy
with the EPR one described above. Then it
is fully within the \Q formalism.\\

It appears that the authors of \cite{EPR}
consider that the paradoxicalness of an
EPR state lies in its contradiction with
completeness of the \QMl description of
an individual bipartite system (which was
claimed by the Copenhagen \i ). Actually,
this {\bf contradiction} may be viewed to
be present in {\bf every entangled
bipartite state} \$\ket{\Psi}_{AB}\$ because it has at least one pair
of twin observables (cf the final parts
of the preceding section). They make
possible selective {\bf distant \m }, and
it creates (or finds) a definite value of
the distant twin observable that was not
a sharp value in \$\ket{\Psi}_{AB}\$.

One can find articles in the literature
in which all entangled bipartite states
are called EPR states. It might be due to
realization of this point. The more so, since Schr\" odinger's view of distant correlations, discussed in the next subsection, brings home this point.\\

Let us return to the singlet state given by (29). (It is hard to find a simpler and better known EPR state.) Let us choose to measure the spin component of the nearby particle B along the z-axis. Let further the \textbf{m}easuring \textbf{i}nstrument be in the initial or ready-to-measure state \$\ket{0}_{mi}\$, and the \textbf{e}xperimenter in the ready-to-watch the result state \$\ket{0}_e\$. The entire four-partite system is in the initial state $$
\ket{\Psi}_{AB}\equiv (1/2)^{1/2}\Big(
\ket{+}_A\ket{-}_B-\ket{-}_A\ket{+}_B\Big)
\otimes\ket{0}_{mi}\ket{0}_e.\eqno{(30)}$$
At the end of the \m , the four-partite system is, e. g., in the state $$\ket{+}_A\ket{-}_B\otimes
\ket{z,-_B,+_A}_{mi}\ket{z,-_B,+_A}_e,
\eqno{(31)}$$ where \$\ket{z,-_B,+_A}_{mi}\$ is the state of the measuring instrument in which the so-called 'pointer position' show the results \$"-"\$ for subsystem B, and \$"+"\$ for the distant subsystem A, and \$\ket{z,-_B,+_A}_e\$ is the analogous state of the experimenter in which the counterpart of the 'pointer position' is the corresponding contents of  consciousness.

Einstein et al. were troubled by the idea that, in transition from (30) to (31), the result \$"+_A"\$ was brought about in a distant action without interaction ( a 'spooky' action), which could not be reconciled with basic physical ideas that reigned outside \qm . It seems to me that the father of relativity ideas in physics has fallen victim to the Bohrian (or Copenhagen) suggestion that (31) describes {\bf absolute reality}. But no wonder; this was more than two decades before Everett's relative-state ideas appeared \cite{Everett1}.

In previous work \cite{FHScully1} (in subsection 7C there) I have adopted, what I call humorously, a 'pocket edition' of Everett's relative-state \I of \qm . (I was sticking to the idea of a laboratory, forgetting about parallel worlds in a multiverse \cite{manyworldsbook}.) I have called the approach {\bf relative reality of unitarily evolving states (by acronym: RRUES)}.

Let me apply RRUES to the above direct \M on subsystem B, and to the simultaneous distant \M on subsystem A.

If the unitary evolution of the system does not change spin projections, then the above initial four-partite state (30) evolves into the state
$$(1/2)^{1/2}\Big(\ket{+}_A\ket{-}_B\otimes
\ket{z,-_B,+_A}_{mi}\ket{z,-_B,+_A}_e\enskip \textbf{+}$$  $$\ket{-}_A\ket{+}_B\otimes
\ket{z,+_B,-_A}_{mi}\ket{z,+_B,-_A}_e\Big).
\eqno{(32)}$$ Here the state (31) is one of the components, one of the 'branches' in Everett's terminology. The point is that the result in (31) is {\bf relative to the state} \$\ket{z,-_B,+_A}_{mi}\ket{z,-_B,+_A}_e\$
of the 'observer'. 'Reality' of the \M results are only relative to the branch in which the 'observer' finds himself. I think this is a suitable realization of Mermin's Ithaca mantra "the correlations, not the correlata" \cite{Mermin}.

One might object that replacing 'absolute reality' of the description of a \Q state by its 'relative reality' is unacceptable. It is well known that for some time the same objection was raised when Einstein replaced absolute motion by relative motion. Nowadays we find no difficulty with it.

Thus, in RRUES there is no 'spooky' action in distance without interaction. One might wonder if 'RRUES' as a new term is justified, when it is pure Everett's relative-state theory. Actually, the new term serves the sole purpose of emphasizing (via the two R's) the {\bf new relativity idea} introduced by Everett in his seminal work.

The outlined interpretation of distant \M by relative-state theory was not published by the present author before.\\

As it was pointed out in Remark 19, any
non-selective or selective \M on the
nearby subsystem B gives rise to distant
state decomposition or distant state
preparation respectively on the distant
subsystem A. One can easily see that two
choices of distinct non-selective direct
\m s on subsystem B can induce state
decompositions on subsystem A that {\it
do not have a common continuation}
(finer decomposition), and hence are
actually incompatible. This might be
viewed as a kind of a {\bf generalized EPR phenomenon}.\\

Realizations of EPR states in thought and real experiments are pointed out in the second and third passage of the Concluding remarks (section 7) below.\\

\subsection{Schr\"odinger's steering}

Relation (12) introduces explicitly the correlation operator into investigations of the effects on the distant subsystem A caused by measurement performed on the nearby  subsystem B . This enabled the Belgrade school to have an original angle and elaborate Schr\" odinger's approach to distant correlations.

The role of the correlation operator in studying distant nearby-subsystem \M effects has thus led to the articles
\cite{steering1}, \cite{steering2}, and \cite{JMPstructure}. But they were written partly in the antlinear Hilbert-Schmidt operators approach, which has been abandoned in this review.\\

For the reader's convenience we rewrite
(and renumerate) relation (12):
$$\ket{\Psi}_{AB}=\sum_{n'}\Big(U_a\rho_B^{1/2}\ket{n'}_B\Big)_A
 \otimes\ket{n'}_B.\eqno{(33a)}$$

Inserting \$U_a^{-1}U_aQ_B\enskip\Big(=Q_B\Big)\$ between \$\rho_B^{1/2}\$ and \$\ket{n'}_B\$ in (33a), which can be done because \$\rho_B^{1/2}Q_B=\rho_B^{1/2}\$, one obtains the equivalent formula
$$\ket{\Psi}_{AB}=\sum_{n'}\rho_A^{1/2}\Big(U_aQ_B
\ket{n'}_B\Big)_A
 \otimes\ket{n'}_B\eqno{(33b)}$$ due to \$U_a\rho_B^{1/2}U_a^{-1}Q_A=\rho_A^{1/2}\$ (cf (14a) etc).

If a nearby-subsystem observable
\$O_B\equiv\sum_{n'}b_{n'}\ket{n'}_B\bra{n'}_B,\enskip
n''\not= n'''\enskip\Rightarrow\enskip b_{n''}\not= b_{n'''}\$, is measured ideally
and selectively having, e. g., the result
\$b_{\bar n}\$ in mind, then
\$\ket{\Psi}_{AB}\$ is hereby converted
into the uncorrelated bipartite state \$\Big(U_a\rho_B^{1/2}\ket{\bar
n}_B\Big)_A\otimes\ket{\bar n}_B\$.
This implies the fact that {\bf the distant
subsystem A is brought into the state} (cf (1f))
$$\ket{\bar n}_A=\Big(U_a\rho_B^{1/2}\ket{\bar
n}_B\Big)_A\Big/
 ||\Big(U_a\rho_B^{1/2}\ket{\bar
 n}_B\Big)_A||=$$
 $$\Big(U_a\rho_B^{1/2}(Q_B\ket{\bar
 n}_B/||Q_B\ket{\bar n}_B||)\Big)_A\Big/
 ||\Big(U_a\rho_B^{1/2}(Q_B\ket{\bar
 n}_B)/||Q_B\ket{\bar
 n}_B||)\Big)_A||.\eqno{(34)}$$ (The fact
 that \$\rho_B^{1/2}=\rho_B^{1/2}Q_B\$ is always valid  was utilized - cf \C 4.)\\

The nearby-subsystem \M that leads to (34) was
called {\bf steering} by Schr\"odinger
\cite{Schroed1}, \cite{Schroed2}, and
'distant steering' in previous work of the present author
\cite{steering1} and \cite{steering2}. It
is also called 'distant preparation' of a
state. It is part of a distant state
decomposition (cf \R 19 above) that is brought
about by the ideal non-selective \M of
the nearby observable \$O_B\$ mentioned above.

Schrodinger pointed out \cite{Schroed1},
\cite{Schroed2} the paradoxical fact that
a skilful experimenter can
steer, without any interaction, a distant
particle (that is correlated with a
nearby one on account of past
interactions) into any of a wide set of
states.\\

The basic steering formula (34) makes clear what the {\bf physical meaning of the correlation operator} \$U_a\$ is. It plays an essential role in determining into which state the distant subsystem is steered. Since this determination takes place jointly with \$\rho_B\$, the physical meaning of \$U_a\$ is much more clear when then action of \$\rho_B\$ is simplified. This is the case when \$\ket{\bar n}_B=\ket{i}_B\$ (cf (15), i. e., when \$\rho_B\ket{\bar n}_B=r_i\ket{\bar n}_B\$. Then \$\rho_B^{1/2}\$ amounts in (34) to multiplication with \$r_i^{1/2}\$, and this has no effect on steering; it affects only the probability (see below). Then \$\ket{\bar n}_B\$ is steered into the state \$\Big(U_a\ket{i}_B\Big)_A=\ket{i}_A\$. If the eigenvalue \$r_i\$ is degenerate, i. e., if \$\cR(Q_B^j)\$ for \$r_j=r_i\$ is at least two dimensional, then the action of \$U_a\$ in mapping  \$\cR(Q_B^j)\$ onto \$\cR(U_aQ_B^jU_a^{-1})\enskip\Big(=\cR(Q_A^j)\Big)\$ (cf the correlated subsystem picture ((14a)-(14d) and (15a-(15f)), is non-trivial. Otherwise, it determines the phase factor of \$\ket{i}_A\$.

Viewing all this in analogy with classical probability theory, one can say that the occurrence of \$\ket{\bar n}_B\bra{\bar n}_B\$ is the condition in the conditional probability, which is the state vector given by the LHS of (34).\\

{\bf Remark 21.} It is obvious from (34)
that all choices of \$\ket{\bar n}_B\$
that have the same projection in
\$\cR(\rho_B)\$ give {\bf the same}
distant state, and if two choices of
nearby state vectors differ only by a
phase factor, so do the corresponding
distant states.\\

{\bf Proposition 5. A)} All states
\$\ket{\phi}_A\$ that belong to
\$\cR(\rho_A^{1/2})\$ and no other states
can be brought about by distant
steering.

{\bf B)} A given state
\$\ket{\phi}_A\in\cR(\rho_A^{1/2})\$ can
be steered into, i. e., it can be given
rise to by selective direct \M of
\$\ket{\bar n}_B\bra{\bar n}_B\$ in
\$\ket{\Psi}_{AB}\$ (cf passage below
(33a)), \IF  $$0\not= Q_B\ket{\bar
n}_B\Big/||Q_B\ket{\bar n}_B||=
\rho_B^{-1/2}U_a^{-1}\ket{\phi}_A\Big/
||\rho_B^{-1/2}U_a^{-1}\ket{\phi}_A||.$$

{\bf Proof. A)} follows immediately from
relation (33b) and B).

{\bf B)} Relation (33a) is seen to imply the claim if one has in mind the fact
that in \$\bar\cR(\rho_B)\$
 \$\rho_B^{1/2}\$ is non-singular and it
 maps \$\bar\cR(\rho_B)\$ onto
 \$\cR(\rho_B^{1/2})\$ in a one-to-one
 way (cf (38) below).\hfill $\Box$\\

\P 5B) implies the lemma of Hadjisavvas \cite{Hadji}: For any given density operator \$\rho\$ a state vector \$\ket{\phi}\$ can appear in a decomposition \$\rho =w\ket{\phi}\bra{\phi}+\sum_kw_k\rho_k\$
(\$W+\sum_kw_k=1\$, each \$\rho_k\$ a density operator, the sum is finite or countably infinite) \IF \$\ket{\phi}\in\rho^{1/2}\$ (let us call it suitability).

That every suitable state vector can appear in a decomposition follows from \P 5B) by performing purification transforming by isomorphism \$\rho\$ into \$\rho_A\equiv\tr_B\Big(\ket{\Psi}_{AB}\bra{\Psi}_{AB}|$
in any way (cf Theorem 2 above), and then taking a basis in \$\cH_B\$ that contains the final state vector in the relation in \P 5B). Clearly, this will give a pure-state decomposition of \$\rho_A\$ in which $\ket{\phi}\bra{\phi}\$ will appear.

That no state vector outside \$\rho^{1/2}\$ can appear in a decomposition can be seen by writing down such a decomposition, then by using it for purification (cf Theorem 2 above), and be getting into contradiction with \P 5.\\

{\bf Remark 22.} A well-known special
case of steering is {\bf erasure}
\cite{erasure}. For instance, the
well-known two-slit interference
disappears when linear polarizers, a
vertical and a horizontal one, are put on
the respective slits \cite{polariz}
because entanglement with the
polarization (internal degree of freedom)
suppresses the coherence. But a \$45^0\$
polarization analyzer can restore (or
revive) the interference. (The
suppressing entanglement is erased.) Here
choice of the analyzer is actually choice
of the state \$\ket{\bar n}_B\$ in
Proposition 5B).\\

One should note that steering is not a
deterministic operation. As it follows
from (33a), the state (34) comes about
with the probability \$p(b_{\bar
n})=||\rho_B^{1/2}\ket{\bar n}_B||^2\$
(because a unitary operator does not
change the norm). As easily seen, one actually has
$$p(b_{\bar n})=||Q_B\ket{\bar
n}_B||^2\times
||\rho_B^{1/2}\Big(Q_B\ket{\bar
n}_B\Big/||Q_B\ket{\bar
n}_B||\Big)||^2.\eqno{(35a)}$$

Relation (35a) implies that all choices of \$\ket{\bar n}_B\$ the
projections in \$\cR(\rho_B)\$ of which
differ only by a phase factor have
the same probability.\\

Since, on account of the positive-eigenvalue eigen-subspaces \$\cR(Q_B^j)\$ of \$\rho_B\$, one has \$Q_B=\sum_jQ_B^j\$, and (35a) can be further rewritten as
$$p(b_{\bar n})=||Q_B\ket{\bar
n}_B||^2\times\sum_j\Big( r_j\times
||\Big(Q_B^j\ket{\bar
n}_B\Big/||Q_B\ket{\bar
n}_B||\Big)||^2\Big).\eqno{(35b)}$$

{\bf Remark 23.} One can see in (35b) that the probability of successful steering (occurrence of \$\ket{\bar n}_B\bra{\bar n}_B\$) is {\bf the larger (i)} if \$\ket{\bar n}_B\$ has a larger projection in the range \$\cR(\rho_B)\$ (if it is 'more' in the range than in the null space), and {\bf (ii)} if the projection is more favorably positioned in the range (if it 'grabs' larger eigenvalues \$r_j\$).\\

On account of \R 23(i), it is practical
to restrict oneself to state vectors from
the range
$$\ket{n}_B=Q_B\ket{n}_B.\eqno{(36)}$$
Choice (36) implies
$$p(b_{n})=||\rho_B^{1/2}\ket{n}_B||^2=
\sum_jr_j\times ||Q_B^j\ket{n}_B||^2\eqno{(37)}$$ (cf (15b)).

In a previous article of the present author \cite{steering2}
Lemmata 1-3 give a detailed mathematical
account of the fine structure of
\$\cR(\rho_B)\$ concerning the action of
\$\rho_B^{1/2}\$. Neither the approach of
writing bipartite state vectors in terms
of antilinear Hilbert-Schmidt operators
that is adopted in the article nor the
results of Lemmata 1-3 do I consider
physically sufficiently important (at the time of writing this
review). Hence it is not reproduced here.
All that should be pointed out is that
one always has
$$\cR(\rho)\subseteq\cR(\rho^{1/2})\subseteq
\bar\cR(\rho),\eqno{(38)}$$ and if
\$dim(\cR(\rho)<\infty\$, then one has
equality throughout in (38), and if {\bf
\$dim(\cR(\rho)=\infty\$}, then both
inclusion relations are {\bf proper}. (It
is also worth pointing out that the mentioned Lemmata 1-3, unlike the rest of the article,
are stated and proved in terms of
standard \QMl arguments.)\\

{\bf Remark 24.} In case of
infinite-dimensional range
\$\cR(\rho_B)\$, the distant states in
\$\bar\cR(\rho_A)\ominus\cR(\rho_A^{1/2})\$,
where \$\ominus\$ denotes set-theoretical
substraction (of a subset), are a kind of {\bf
irrationals} concerning steering: one
cannot steer the distant subsystem into
these states exactly, but one can achieve
this arbitrarily closely (because
\$\cR(\rho_A^{1/2})\$ is dense in
\$\bar\cR(\rho_A)\$, cf (38)).\\

{\bf Remark 25.} As it was pointed out in
Remark 19, one can perform \M of an
incomplete observable \$O_B\$, i. e., one
that has degenerate eigenvalues, on the
nearby subsystem and obtain distant state
decomposition in the non-selective
version, or state preparation in the
selective version. In the latter case one
has {\bf generalized steering}, which
results, in general, in a mixed state of
the distant subsystem.\\

Schr\"odinger's steering has recently
drawn much attention. For example,
steering was generalized to mixed states
in \cite{steering.mixed}. Asymmetric
steering was studied in
\cite{steer.asym}. (See also the review
article in \cite{review}.)\\

\section{Concluding remarks}

Under the title "On bipartite pure-state
entanglement structure in terms
of disentanglement" in
\cite{JMPstructure} Schr\" odinger's
disentanglement, i.e., distant state
decomposition, as a physical way to study
entanglement, is carried one step further
with respect to previous work in
investigating the
qualitative side of entanglement in any
bipartite state vector. Distant
measurement or, equivalently, distant
orthogonal state decomposition from
previous work (cf \R 17  and \R 18 above) is
generalized to distant linearly
independent complete state decomposition
both in
the non-selective and the selective
versions (cf \R 19 above). The results
are displayed in terms of commutative
square diagrams, which show the power and
beauty of the physical meaning of the
antiunitary
correlation operator \$U_a\$ inherent in
any given bipartite state vector
\$\ket{\Psi}_{AB}\$. It is shown that
linearly independent distant pure-state
preparation, which is caused by selective
\M  of an observable \$O_B\$ on the
nearby system that does not commute with
its state operator \$\rho_B\$ (cf \T 6
above), carries the {\bf highest
probability of occurrence} among  distant
preparations that are not obtained by
selective distant \m .\\

Under the titles "On EPR-Type
Entanglement in the Experiments
of Scully et al. I. The Micromaser Case
and Delayed-Choice Quantum Erasure" and
"On EPR-type Entanglement in the
Experiments of Scully et al. II. Insight
in the Real Random Delayed-choice Erasure
Experiment" in \cite{FHScully1} and
\cite{FHScully2} respectively intricate
realizations of EPR states in a thought
experiment and a real experiment
respectively are discussed.\\

In the yet unpublished preprint under the
title "Quantum Correlations in
Multipartite States. Study Based on the
Wootters-Mermin Theorem"
\cite{FHArxivMermin} a nice example of an
EPR state is given in relation (15) in
section 7 there.\\

In the article \cite{twincoh} under
the title "The role of coherence entropy
of physical twin observables in
entanglement" the concept of twin
observables for bipartite quantum states
is simplified. The relation of observable
and state is studied in detail from the
point of view of coherence entropy.\\

In the article \cite{Pauli} under the
title "Irrelevance of the Pauli principle
in distant correlations
between identical fermions" it was shown
that the Pauli non-local correlations do
not contribute to distant correlations
between identical fermions. In distant
correlations a central role is played by
distant \M (cf subsection 6.1 above). A
negentropy measure of distant
correlations is introduced and discussed.
It is demonstrated that distant
correlations are
necessarily of dynamical origion.\\

In the short article \cite{constr.all}
under the title "How to define
systematically all possible two-particle
state vectors in terms of conditional
probabilities" all bipartite state
vectors of given subsystems were
systematically generated using the state
operator \$\rho_B\$ of the nearby
subsystem and the correlation operator
\$U_a\$ (cf sections 3. and 4. above).\\

Under the title "Complete "Born's rule"
from "environment-assisted invariance"
in terms of pure-state twin unitaries" in
\cite{envariance1} the concept of twin
observables was extended to twin
unitaries. It was shown that the latter
are the other face of Zurek's envariance
concept.\\

Under the title "Mixed-state twin
observables" in \cite{twin.mixed} the
twin-observables notion was extended to
bipartite mixed states (density
operators) \$\rho_{AB}\$. It was shown
that commutation of the twin observables
with the corresponding state operators
\$[O_A,\rho_A]=0\$ and \$[O_B,\rho_B]=0\$
are necessary conditions also for mixed
states, but these relations are no
longer sufficient.\\

Under the title "Hermitian Schmidt
decomposition and twin observables of
bipartite mixed states" in
\cite{Schmidt.mixed} It was shown that
every mixed bipartite state (density
operator) \$\rho_{AB}\$ has a Schmidt
decomposition in terms of Hermitian
subsystem operators. This result is due
to the fact that \$\rho_{AB}\$ is an
element in the Hilbert space of all
linear Hilbert-Schmidt operators in
\$\cH_A\otimes\cH_B\$.\\

In the article under the title "On
statistical and deterministic quantum
teleportation" in \cite{telep} it was
shown that use of correlation operators
gives insight in teleportation (cf Figure
2 in section 6 there).\\

In the preprint under the title "Delayed
Twin Observables Are They a Fundamental
Concept in Quantum Mechanics?" in
\cite{twins.delayed} the twin-observables
concept is generalized to the case when
unitary time evolution takes place.\\

Finally, it is worth reemphasizing
that all results presented in sections
2-6 apply to {\bf every} bipartite state
vector. For instance, in \$\ket{\Psi}_{AB}\$ subsystem
A can be the orbital, and subsystem B the
spin degree of freedom of one electron,
but it can also describe a
many-particle system in which A contains
some of the particles and B contains the
rest.\\

The correlation operator provides us with
a way to comprehend entanglement in a
bipartite pure state. It primarily serves
to give insight. For most practical
purposes the canonical Schmidt
decomposition or its stronger form, a
twin-adapted canonical Schmidt
decomposition, suffice. The correlation
operator is implicit in it.\\

The elaborated systematic and comprehensive analysis presented should, hopefully,
enable researchers to utilize Schmidt
decomposition as a scalpel in surgery to
derive new results. At least I was myself enabled by it to work out a detailed
theory of exact \QMl \m , which will be
presented elsewhere.\\

{\bf\noindent Appendix A. Partial scalar
product}\\

It will be shown that partial scalar
product can be defined in three and a 'half' ways, i. e., in three equivalent ways and incompletely in a fourth way.

We still write arbitrary ket or bra vectors
with a bar; those without a bar are norm-one vectors (as it is in the text). In each of the definitions below, we define the partial scalar product only for norm-one elements of the Hilbert spaces. If the norm of any (or both) of the factors in the product is not one, the final element is, by part of the definition,  multiplied by this norm (or by both norms).\\

{\bf A)} {\it Definition in terms of subsystem-basis expansion.} We define partial scalar product by essentially  equating RHS(1g) and RHS(1b). More precisely, for any norm-one element \$\ket{n}_B\enskip\Big(\in\cH_B\Big)\$ and any norm-one element \$\ket{\Psi}_{AB}\enskip
\Big(\in(\cH_A\otimes\cH_B)\Big)\$
we write:
$$\Big(\bra{n}_B\ket{\Psi}_{AB}\Big)_A
\equiv  \sum_m(\bra{m}_A\bra{n}_B
\ket{\Psi}_{AB})\times\ket{m}_A.\eqno{(A.1)}$$ (Note that the resulting element in \$\cH_A\$ is expanded in an arbitrary basis \$\{\ket{m}_A:\forall m\}\$.)

Next, we derive {\it two basic properties} of partial scalar product from the definition.

{\it Property (i).} If the bipartite element is {\it uncorrelated} \$\ket{\Psi}_{AB}=\ket{\psi}_A\otimes
\ket{\phi}_B\$, then partial scalar product {\it reduces to ordinary scalar product}: $$\Big(\bra{n}_B(\ket{\psi}_A\otimes
\ket{\phi}_B)\Big)_A=
(\bra{n}_B\ket{\phi}_B)\times\ket{\psi}_A.\eqno{(A.2)}$$
This obviously follows from (A.1).

{\it Property (ii).} If the bipartite element is {\it expanded in an absolutely convergent orthogonal series} \$\ket{\Psi}_{AB}=
\sum_k\overline{\ket{\Psi}}_{AB}^k\$ (it can be a double etc. series), then the partial scalar product has the property of {\it extended linearity}:
$$\Big(\bra{n}_B(\sum_k\overline{
\ket{\Psi}}_{AB}^k)\Big)_A=
\sum_k\Big(\bra{n}_B\overline{\ket{\Psi}}_{AB}^k\Big)_A.
\eqno{(A.3)}$$ Also (A.3) follows evidently from (A.1) if one takes into account the fact that two absolutely converging series (or double series etc.) can exchange order.\\

One can evaluate the form of the partial scalar product in the representation of arbitrary bases \$\{\ket{m}_A:\forall m\}\$ in \$\cH_A\$ and \$\{\ket{q}_B:\forall q\}\$ in \$\cH_B\$: $$\Big(\bra{n}_B\ket{\Psi}_{AB}\Big)_A\equiv\sum_m(\bra{m}_A
\bra{n}_B
\ket{\Psi}_{AB})\times\ket{m}_A=\sum_m\Big[\bra{m}_A\bra{n}_B
\Big(\sum_q\ket{q}_B\bra{q}_B\Big)\ket{\Psi}_{AB}\Big]
\times\ket{m}_A=$$
$$\sum_m\sum_q(\bra{n}_B
\ket{q}_B)\times ( \bra{m}_A\bra{q}_B\ket{\Psi}_{AB})\times\ket{m}_A=
\sum_m\Big(\sum_q(\bra{n}_B\ket{q}_B)\times\bra{m}_A \bra{q}_B\ket{\Psi}_{AB}\Big)\times\ket{m}_A.$$

Thus, partial scalar product in the representation in the basis \$\{\ket{q}_B:\forall q\}\$ (the q-representation) is $$\bra{m}_A\Big(\bra{n}_B\ket{\Psi}_{AB}\Big)_A=
\sum_q[(\bra{q}_B\ket{n}_B)^*\times (\bra{m}_A \bra{q}_B\ket{\Psi}_{AB})],\eqno{(A.4)}$$ where the asterisk denotes complex conjugation.

The q-representation can be also purely continuous (as the coordinate or linear momentum representations). Then (A.4) has the form $$\bra{m}_A\Big(\bra{n}_B\ket{\Psi}_{AB}\Big)_A=
\int_q[(\bra{q}_B\ket{n}_B)^*\times (\bra{m}_A \bra{q}_B\ket{\Psi}_{AB})].\eqno{(A.5)}$$\\

{\bf B)} {\it Definition in terms of properties (i) and (ii).} If we assume the validity of the two basic properties from above, then, substituting the suitable general expansion (1e) for \$\ket{\Psi}_{AB}\$ in \$\Big(\bra{n}_B\ket{\Psi}_{AB}\Big)_A\$ one recovers (1b), and one is back to the subsystem-basis-expansion definition (A) above. Therefore, definitions (A) and (B) are {\it equivalent}.\\

{\bf C)} {\it Definition of the partial scalar product in representation.} We define the partial scalar product by (A.4). Reading the above derivation of (A.4) backwards, we recover the sub-system-basis-expansion definition (A). Hence, definitions (A) and (C) are equivalent.\\

{\bf D)} {\it Definition of the partial scalar product in terms of the partial trace up to a phase factor} is given in Proposition C.1 in Appendix C below.\\

{\it Remark A.1} As easily seen,  the partial scalar product \$\overline{\bra{\phi}}_B
\overline{\ket{\Psi}}_{AB}\$ ca be
evaluated also by expressing
\$\overline{\ket{\Psi}}_{AB}\$ as any (finite)  linear
combination  of tensor products of
tensor-factor vectors.\\

{\bf\noindent Appendix B. The
partial-trace and its rules.}\\

The partial trace $$\bra{m}_A\tr_BO_{AB}\ket{m'}_A\equiv\sum_n
\bra{m}_A\bra{n}_B\rho_{AB}\ket{m'}_A\ket{n}\eqno{(B.1)}$$
was explained in von Neumann's book \cite{vNeum} (p. 425) as far as \$O_{AB}\equiv\rho_{AB}\$, a composite-system density operator was concerned. The so-called 'reduced' entity (on the LHS) is defined by (B.1) in bases in an apparently basis-dependent way. But the resulting positive operator \$\rho_A\enskip\Big(\equiv\tr_B\rho_{AB}\Big)\$ of finite trace is basis independent.

The very concept of a partial trace comes from the fact that one can have a {\it state operator} (density operator; generalization of state vector) describing a subsystem as follows. For every first-subsystem observable \$O_A\otimes I_B\$ one obtains $$\Big<O_A,\rho_{AB}\Big>=\tr\Big(\rho_{AB}
(O_A\otimes I_B)\Big)=\tr_A\Big[
\Big(\tr_B\rho_{AB}\Big)O_A\Big]=\tr_A\Big(\rho_A
O_A\Big).\eqno{(B.2)}$$ The second partial-trace rule (cf below) has been used. (Note that the in full trace "$\tr=\tr_A\tr_B$" the indices are usually omitted as superfluous.)\\

{\bf FIRST RULE} ({\it The 'commutation-under-the-partial-trace' rule.}) If \$\cH_A\otimes\cH_B\$
is a two-subsystem (complex and
separable) composite Hilbert space, if,
further, \$O_A\$ is an operator that acts
non-trivially only in \$\cH_A\$ and
\$O_{AB}\$ is any operator in the
composite Hilbert space, then the
following partial-trace rule is valid
$$\tr_A\Big(O_AO_{AB}\Big)\enskip\textbf{=}\enskip
\tr_A\Big(O_{AB}O_A\Big).\eqno{(B.3)}$$
(Naturally, \$O_A\$ is actually
\$O_A\otimes I_B\$ when acting in
\$\cH_A\otimes\cH_B\$.)

Symmetrically,
$$\tr_B\Big(O_BO_{AB}\Big)\enskip\textbf{=}\enskip
\tr_B\Big(O_{AB}O_B\Big).\eqno{(B.4)}$$
Rules (B.3) and (B.4) are analogous to
commutation under a full trace.\\

{\bf Proof.} Let \$\{\ket{r}_A:\forall
r\}\$ and \$\{\ket{s}_B:\forall s\}\$ be
any complete ON bases in the
factor spaces. Then, in view of
\$\bra{s}_BI_B\ket{s'}_B=\delta_{s.s'}\$,
one can write
$$\bra{s}_BLHS(B.3)\ket{\bar
s}_B=\sum_{r'r''s'}\bra{r'}_A\bra{s}_B(O_A\otimes
I_B)\ket{r''}_A\ket{s'}_B\bra{r''}_A\bra{s'}_BO_{AB}
\ket{r'}_A\ket{\bar s}_B=$$
$$\sum_{r'r''}\bra{r'}_AO_A\ket{r''}_A
\bra{r''}_A\bra{s}_BO_{AB}\ket{r'}_A\ket{\bar
s}_B.\eqno{(B.5)}$$

On the other hand,
$$\bra{s}_BRHS(B.3)\ket{\bar
s}_B=\sum_{r'r''s'}\bra{r'}_A\bra{s}_B
O_{AB}\ket{r''}_A\ket{s'}_B\bra{r''}_A\bra{s'}_B
(O_A\otimes I_B)\ket{r'}_A\ket{\bar
s}_B=$$
$$\sum_{r'r''}\bra{r'}_A\bra{s}_B
O_{AB}\ket{r''}_A\ket{\bar s}_B
\bra{r''}_AO_A\ket{r'}_A.\eqno{(B.6)}$$

If one exchanges the order of the two
(number) factors and also exchanges the
two mute indices \$r'\$ and \$r''\$ in
each term on the RHS of (B.6), then the
RHS's of (B.5) and (B.6) are seen to be
equal. Hence, so are the LHS's. Rule
(B.4) is proved analogously.\hfill
$\Box$\\

{\bf SECOND RULE} ({\it The 'out-of-the-partial-trace' rule.}) Under the assumptions
of the first rule, the following
relations are always valid:
$$\tr_B\Big(O_AO_{AB}\big)=O_A\tr_BO_{AB}.\eqno{(B.7)}$$
$$\tr_B\Big(O_{AB}O_A\big)=(\tr_BO_{AB})O_A.\eqno{(B.8)}$$
$$\tr_A\Big(O_BO_{AB}\big)=O_B\tr_AO_{AB}.\eqno{(B.9)}$$
$$\tr_A\Big(O_{AB}O_B\big)=(\tr_BO_{AB})O_B.\eqno{(B.10)}$$
An operator that acts non-trivially only
in the tensor-factor space that is
opposite to the one over which the
partial trace is taken behaves
analogously as a constant under a full
trace: it can be taken outside the
partial trace. But one must observe the
order (important for operators, not for
numbers).\\

{\bf Proof.} Let \$\{\ket{r}_A:\forall
r\}\$ and \$\{\ket{s}_B:\forall s\}\$ be
any complete ON bases in the
factor spaces. Then
$$\bra{r}_ALHS(B.7)\ket{r'}_A=\sum_{r''ss'}
\bra{r}_A\bra{s}_B(O_A\times
I_B)\ket{r''}_A\ket{s'}_B
\bra{r''}_A\bra{s'}_BO_{AB}\ket{r'}_A\ket{s}_B=$$
$$\sum_{r''s}\bra{r}_AO_A\ket{r''}_A
\bra{r''}_A\bra{s}_BO_{AB}\ket{r'}_A\ket{s}_B.\eqno{(B.11)}$$

On the other hand,
$$\bra{r}_ARHS(B.7)\ket{r'}_A=\sum_{r''s}
\bra{r}_AO_A\ket{r''}_A
\bra{r''}_A\bra{s}_BO_{AB}\ket{r'}_A\ket{s}_B.\eqno{(B.12)}$$
The RHS's of (B.11) and (B.12) are seen to
be equal. Hence, so are the LHS's.
Relations (B.8), (B.9) and (B.10) are
proved analogously.\hfill $\Box$\\

{\bf\noindent Appendix C. Equivalence of the partial scalar product and a certain partial trace.}\\

The auxiliary relations that follow stand in certain analogies with the known basic relation $$ \tr(\ket{\Psi}_{AB}\bra{\Psi}_{AB}O_{AB})=
\bra{\Psi}_{AB}O_{AB}\ket{\Psi}_{AB}\eqno{(C.1)}$$
(obvious if one evaluates the trace in a basis in which \$\ket{\Psi}_{AB}\$ is one of the elements).

{\bf Lemma C.1}
$$\tr_B\Big((\ket{\phi}_B\bra{\phi}_B) (\ket{\Psi}_{AB}\bra{\Psi}_{AB})\Big)=
\overline{\bra{\phi}_B\ket{\Psi}_{AB}}\:
\overline{\bra{\Psi}_{AB})\ket{\phi}_B}.$$\\

{\bf Proof.} Utilizing definition (B.1), taking into account that \$\bra{m}_AI_A\ket{\bar m}_A=\delta_{m,\bar m}\$, and eventually making use of (1b),one obtains $$\bra{m}_ALHS\ket{m'}_A=\sum_n\bra{m}_A\bra{n}_B
\Big((\ket{\phi}_B\bra{\phi}_B) (\ket{\Psi}_{AB}\bra{\Psi}_{AB})\Big)\ket{m'}_A\ket{n}_B=$$
$$\sum_{n,\bar m,\bar n}\bra{m}_A\bra{n}_B(I_A\otimes\ket{\phi}_B\bra{\phi}_B)
\ket{\bar m}_A\ket{\bar n}_B\bra{\bar m}_A\bra{\bar n}_B
(\ket{\Psi}_{AB}\bra{\Psi}_{AB})\ket{m'}_A\ket{n}_B=$$
$$\sum_{n,\bar n}\bra{n}_B\ket{\phi}_B\times\bra{\phi}_B)\ket{\bar n}_B\times\bra{m}_A\bra{\bar n}_B\ket{\Psi}_{AB}\times\bra{\Psi}_{AB})\ket{m'}_A\ket{n}_B=$$
$$\Big(\sum_{\bar n}\bra{\phi}_B\ket{\bar n}_B\times\bra{m}_A\bra{\bar n}_B\ket{\Psi}_{AB}\Big)\times\Big(\sum_n \bra{\Psi}_{AB}\ket{m'}_A\ket{n}_B\bra{n}_B
\ket{\phi}_B\Big)=$$
$$\bra{m}_A\bra{\phi}_B\ket{\Psi}_{AB}\times
\bra{\Psi}_{AB}\ket{m'}_A\ket{\phi}_B=$$
$$\bra{m}_ARHS\ket{m'}_A.$$\hfill $\Box$\\

{\bf Lemma C.2}
$$\tr\Big((\ket{\phi}_B\bra{\phi}_B)
(\ket{\Psi}_{AB}\bra{\Psi}_{AB})\Big)= ||\overline{\bra{\phi}_B\ket{\Psi}_{AB}}
||^2$$

{\bf Proof.} According to Lemma C.1
$$LHS=\tr\Big(\overline{\bra{\phi}_B\ket{\Psi}_{AB}}\:
\overline{\bra{\Psi}_{AB})\ket{\phi}_B}\Big)=$$
$$||\overline{\bra{\phi}_B\ket{\Psi}_{AB}}||\times\Big\{
\tr\Big[\Big(\overline{\bra{\phi}_B\ket{\Psi}_{AB}}\Big/
||\overline{\bra{\phi}_B\ket{\Psi}_{AB}}||\Big)
\Big(\overline{\bra{\Psi}_{AB})\ket{\phi}_B}\Big/
||\overline{\bra{\Psi}_{AB})\ket{\phi}_B}||\Big]\Big\}\times$$ $$||\overline{\bra{\Psi}_{AB})\ket{\phi}_B}||=RHS$$\hfill $\Box$

Finally, the two lemmata obviously imply the claim:

{\bf Proposition C.1 } The following bridge relation is valid between partial trace and partial scalar product:
$$\tr_B\Big(\ket{\phi}_B\bra{\phi}_B)
(\ket{\Psi}_{AB}\bra{\Psi}_{AB})\Big)
 \Big/\Big[\tr\Big(\ket{\phi}_B\bra{\phi}_B)
 (\ket{\Psi}_{AB}\bra{\Psi}_{AB})\Big)\Big]=$$  $$\Big(\overline{\bra{\phi}_B\ket{\Psi}_{AB}}\Big/
||\overline{\bra{\phi}_B\ket{\Psi}_{AB}}||\Big) \:\Big(\overline{\bra{\Psi}_{AB})\ket{\phi}_B}\Big/
||\overline{\bra{\Psi}_{AB})\ket{\phi}_B}||\Big).$$

\end{document}